\documentclass[aps,pra,twocolumn,10pt,groupedaddress]{revtex4-2}

\usepackage[utf8]{inputenc}
\usepackage{amsmath}
\usepackage{amssymb}
\usepackage{graphicx}
\usepackage[percent]{overpic}
\usepackage{enumerate}
\usepackage{mathbbol}
\usepackage{xspace}
\usepackage{hyperref}
\usepackage[section]{placeins}

\usepackage{xcolor}
\hypersetup{
    colorlinks,
    linkcolor={red!85!black},
    citecolor={blue!85!black},
    urlcolor={blue!85!black}
}

\newcommand{\widesim}[2][1.5]{
  \mathrel{\underset{#2}{\scalebox{#1}[1]{$\sim$}}}
}

\newcommand{\basis}{MPB\xspace}
\newcommand{\purity}{R}
\newcommand{\dpuritydt}{\dot{R}}

\definecolor{orange}{RGB}{223,101,0}

\global\long\def\CG#1#2#3#4#5#6{C_{#2\, #4\, #6}^{#1\, #3\, #5}}

\global\long\def\ket#1{|#1\rangle}
\global\long\def\n{\mathbf{n}}
\global\long\def\Jn{\mathbf{J}\boldsymbol{\cdot}\mathbf{n}}

\global\long\def\bra#1{\langle#1|}

\global\long\def\tr{\operatorname{Tr}}
\global\long\def\HOAP{\textrm{HOAP}\xspace}

\begin{document}

\title{Extreme depolarization for any spin}

\author{Jérôme Denis}
\email{jdenis@uliege.be}
\affiliation{Institut de Physique Nucléaire, Atomique et de Spectroscopie, CESAM, University of Liège, B-4000 Liège, Belgium}

\author{John Martin}
\email{jmartin@uliege.be}
\affiliation{Institut de Physique Nucléaire, Atomique et de Spectroscopie, CESAM, University of Liège, B-4000 Liège, Belgium}

\date{February 15, 2022}

\begin{abstract}
The opportunity to build quantum technologies operating with elementary quantum systems with more than two levels is now increasingly being examined, not least because of the availability of such systems in the laboratory. It is therefore essential to understand how these single systems initially in highly non-classical states decohere on different time scales due to their coupling with the environment. In this work, we consider the depolarization, both isotropic and anisotropic, of a quantum spin of arbitrary spin quantum number~$j$ and focus on the study of the most superdecoherent states. We approach this problem from the perspective of the collective dynamics of a system of $N=2j$ constituent spin-$1/2$, initially in a symmetric state, undergoing collective depolarization. This allows us to use the powerful language of quantum information theory to analyze the fading of quantum properties of spin states caused by depolarization. In this framework, we establish a precise link between superdecoherence and entanglement. We present extensive numerical results on the scaling of the entanglement survival time with the Hilbert space dimension for collective depolarization. We also highlight the specific role played by anticoherent spin states and show how their Markovian isotropic depolarization alone can lead to the generation of bound entangled states.
\end{abstract}

\maketitle

\section{Introduction} \label{sec:intro}

The elementary building blocks for storing and processing information on quantum devices are quantum bits or qubits. Qubits are controllable two-level quantum systems that can take very different forms, from the spin of an electron in NV center, to trapped ions, neutral atoms in optical lattices, photons, or superconducting circuits. A key figure of merit for the quality of a qubit is its coherence time,  i.e., the time during which it can reliably store information or remain in a quantum superposition of states before decoherence transforms it into a statistical mixture of states.
Efforts to use elementary quantum systems with more than two degrees of freedom for quantum technologies have been intensified by the availability of such systems in experiments operating in the quantum regime. These so-called qudits can be implemented with a variety of systems, e.g., using the orbital angular momentum of light~\cite{2018Zeilinger}, the electronic spin of magnetic atoms~\cite{2021Satoor}, the internal levels of trapped ions~\cite{Ionqudits}, superconducting circuits~\cite{SCqudits}, or the rotational degree of freedom in molecules~\cite{Saw20}.
This work explores previously unaddressed fundamental questions related to the decoherence of these individual qudits or, equivalently, of individual spins of arbitrary quantum number $j$. Most of the work done so far on the depolarization of spin states has either followed a phenomenological approach based on a SU($n$) depolarization channel~\cite{2002Kempe,2004Dur,2005Dur,2005Lidar,Aol08,2010Nori}, or has focused on dynamical decoherence of specific states, such as NOON states and coherent states~\cite{2013Rivas,Benedict99}. Our work is different in its focus and in the spin states that play a central role: the anticoherent spin states. These states have recently attracted much attention because of their usefulness in quantum metrology~\cite{Gol18, Mar20, Gol21}. Here, we aim to provide additional insight on the dynamical depolarization of spin states by addressing the following questions: Which states are most prone to decoherence on short-time scales but also on longer time scales (states reaching a quantum speed limit) for any spin ? What is the quantum information theoretic resource that enables superdecoherence ? How long does it take before a spin state becomes absolutely classical due to decoherence, in the sense that non-classicality of the state cannot be revived by the sole application of unitary transformations~?

More precisely, in this work, we study depolarization under strict conservation of energy between the spin and its environment~\cite{2013Kosloff, 2020Kosloff}. This condition, which stems from thermodynamic considerations and applies to every microscopic model of decoherence that leads to a Markovian master equation, induces constraints on decoherence rates. The problem of the depolarization of a single spin $j$ can be mapped onto the collective decoherence of $N=2j$ spin $1/2$ initially in a symmetric state. As we shall see, the time evolution has two important properties: it is unital and it preserves separability. The general dynamics is then as follows (for depolarization in at least two directions): An initial pure state $\ket{\psi_0}$ evolves into a mixed state $\rho$ whose purity
\begin{equation}\label{purity}
R(\rho)\equiv\tr[\rho^2]
\end{equation}
decreases monotonically with time as $\rho$ converges to the unique stationary state, the maximally mixed state (MMS)
\begin{equation}\label{MMS}
\rho_0 \equiv \frac{\mathbb{1}}{2j+1},
\end{equation}
proportional to the identity operator $\mathbb{1}$ on $\mathcal{H}\simeq \mathbb{C}^{2j+1}$. In the course of this evolution, the entanglement initially present, which we quantify by the negativity across different bipartitions of the $N$ spin-$1/2$ system, decreases and ends up vanishing, leading on the way to bound-entangled states with respect to certain bipartitions. Once the state becomes separable, which occurs after a finite time, it remains separable at any subsequent time. The state eventually enters the ball of absolutely separable states and no entanglement can be recovered by unitary evolution alone. We want to study how fast quantum correlations decrease over time, which pure states decohere the most rapidly or how long it takes before the state enters a ball of absolutely separable states. This work thus intends to deepen our understanding of the dynamics of decoherence associated with collective depolarization processes.

This paper is organized as follows. In Sec.~\ref{Sec:states}, we introduce the tools needed for our work. In Sec.~\ref{Sec:CollectiveDecoherence}, we present the Lindblad master equation describing the depolarization of an arbitrary spin and its expression in the multipole operator basis. We discuss a conservation law for the dynamics, its steady states, and a general solution for the density matrix and its reductions. In Sec.~\ref{Sec:extrisotrop}, we study isotropic depolarization, from the evolution of purities, the condition of appearance of superdecoherence, the identification of the states most sensitive  to decoherence to the dynamics of entanglement. Finally, in Sec.~\ref{Sec:anisotrop}, we study the same type of questions for anisotropic depolarization. Most of the technical derivations of this work are relegated to the Appendixes. 

\section{Spin states, classicality, anticoherence, and entanglement}\label{Sec:states}

In this section, we introduce the concepts and tools about spin states useful for this work, such as some of their representations (Sec.~\ref{spinStates}), the notions of classicality and absolute classicality (Sec.~\ref{spinStatesClassicality}), the concept of anticoherence (Sec.~\ref{subsecAC}), and the isomorphism between spin states and symmetric multiqubit states (Sec.~\ref{SymmMultiQubit}) enabling us to use the tools of quantum information theory.

\subsection{Representations of spin states}\label{spinStates}
Let $\rho$ be a density operator describing the state of a spin-$j$ quantum system. In this work, we will use several different representations for $\rho$.

\paragraph{Angular momentum basis.} A first representation of $\rho$ follows from its expansion in the eigenbasis of the operators $\hat{\mathbf{J}}^2,\hat{J}_z$. The general form of this expansion reads as
\begin{equation}\label{rhosb}
\rho=\sum_{m,m'=-j}^j\rho_{mm'}|j,m\rangle\langle j,m'|,
\end{equation}
where $|j,m\rangle$ are the standard angular momentum basis states satisfying
\begin{equation}\label{standardangularbasis}
\begin{aligned}
& \hat{\mathbf{J}}^2|j,m\rangle=j(j+1)\hbar^2|j,m\rangle,\\
& \hat{J}_z|j,m\rangle=m\hbar|j,m\rangle,
\end{aligned}
\end{equation}
for $m=-j,\ldots,j$. For a pure state such as a spin-coherent state $\ket{\n}$, which is defined as the eigenstate of the angular momentum component $\Jn$ along the unit vector $\mathbf{n}\in\mathbb{R}^3$ with maximal eigenvalue $\hbar j$, the expansion takes the form
\begin{equation}\label{cohjm}
\ket{\n}=\!\sum_{m=-j}^j \!\sqrt{\tbinom{2j}{j-m}}[\cos(\tfrac{\theta}{2})]^{j+m} [\sin(\tfrac{\theta}{2})e^{i\varphi}]^{j-m} \ket{j,m},
\end{equation}
where 
\begin{equation}
\mathbf{n}=(\sin\theta\cos\varphi, \sin\theta\sin\varphi,\cos\theta)^T
\end{equation}
with $\theta\in[0,\pi]$ and $\varphi\in[0,2\pi[$.

\paragraph{Multipole operator basis (\basis).} Alternatively, one can expand the density operator $\rho$ in terms of multipole operators $T_{LM}$ (see Appendix \ref{Appendix_MOB} for a definition and some properties of these operators) as 
\begin{equation}\label{rhoexpTLMs}
\rho=\sum_{L=0}^{2j}\sum_{M=-L}^{L}\rho_{LM}\,T_{LM}.
\end{equation}
For example, in the multipole operator basis, a spin-coherent state has the expansion~\cite{Arecchi72}
\begin{equation}\label{cohTLM}
\ket{\n}\bra{\n}
=\sum_{L=0}^{2j}\sum_{M=-L}^{L}\rho_{LM}(\Omega)\,T_{LM}
\end{equation}
with state multipoles
\begin{equation}\label{rhoLMspinc}
\rho_{LM}(\Omega)= \frac{\sqrt{4\pi}\,(2j)!}{\sqrt{(2j-L)!(2j+L+1)!}}\,Y_{LM}^*(\Omega),
\end{equation}
where $Y_{LM}(\Omega)$ are the spherical harmonics and $\Omega\equiv (\theta,\varphi)$. Equation~\eqref{rhoLMspinc} shows that in the case of spin-coherent states, all the state multipoles from $L=0$ to $L=2j$ are non zero. 

For future use, we note that, in the \basis, the purity \eqref{purity} takes the simple form (see Appendix \ref{Appendix_MOB})
\begin{equation}\label{purityrhoLM}
\purity(\rho)=\sum_{L=0}^{2j}\sum_{M=-L}^{L}|\rho_{LM}|^{2}.
\end{equation}

\paragraph{Overcomplete basis of spin-coherent states.} Another representation, valid for any linear operator acting on $\mathcal{H}$, follows from the fact that spin-coherent states form an overcomplete set from which the identity operator can be expressed as
\begin{equation}\label{overspincoh}
\mathbb{1}=\frac{2j+1}{4 \pi} \int \ket{\n}\bra{\n} \,d \Omega
\end{equation}
with $d \Omega=\sin\theta d\theta d\varphi$. As for bosonic states, spin states admit a diagonal or $P$ representation~\cite{Arecchi72}
\begin{equation}\label{rhoPfuncgen}
\rho=\frac{2j+1}{4 \pi} \int P_{\rho}(\Omega)\,\ket{\n}\bra{\n}\, d \Omega,
\end{equation}
with $P_{\rho}(\Omega)$ the $P$ function of $\rho$. There is, however, one important difference from the bosonic case which is that the $P$ function is not unique, see e.g.~\cite{2017Giraud}.

The link between the last two representations is made by expanding the multipole operators in terms of spin-coherent states as
\begin{equation}\label{TLMspincoh}
T_{LM}=\frac{2j+1}{4 \pi} \int \alpha_{L}^{(2j)}\,Y_{LM}(\Omega) \ket{\n}\bra{\n} \, d \Omega
\end{equation}
with $\alpha_{L}^{(2j)}$ the constant
\begin{equation}
\alpha_{L}^{(2j)}=\frac{2\sqrt{\pi}}{(2j+1)!}\sqrt{(2j-L)!(2j+L+1)!}.
\end{equation}
By inserting Eq.~\eqref{TLMspincoh} into Eq.~\eqref{rhoexpTLMs}, one can deduce the expression of a valid $P$ function for $\rho$ in terms of its state multipoles, which is given by
\begin{equation}\label{PfunctrhoLM}
P_{\rho}(\Omega)=\sum_{L=0}^{2j}\sum_{M=-L}^{L}  \alpha_{L}^{(2j)} \rho_{LM} \,Y_{LM}(\Omega).
\end{equation}

\subsection{Classicality of spin states}\label{spinStatesClassicality}
\paragraph{Classical spin state.} A spin state $\rho$ is said to be \textit{classical} if it can be written as a convex combination of spin-coherent states, i.e., if there exists a function $P_{\rho}(\Omega)\geqslant 0\; \forall\;\Omega$ such that Eq.~\eqref{rhoPfuncgen} holds~\cite{2008Giraud}. Otherwise, it is said to be non-classical. Because the $P$ function is not unique, a classical state may also be described by other $P$ functions taking negative values for certain values of their arguments~\cite{2017Giraud}.

\paragraph{Absolutely classical spin state.} A spin state is said to be \textit{absolutely classical} when it is classical and remains classical under any global unitary transformation $U$ acting on $\mathcal{H}$. Absolutely classical spin states can only be mixed states, an example being given by the MMS $\rho_0$ [see Eq.~\eqref{MMS}].

\paragraph{Distance to the maximally mixed state.} We define the distance $r$ of a state $\rho$ to the MMS by writing
\begin{equation}\label{definition_r}
\rho=\rho_0+r\,\tilde{\rho},
\end{equation}
where $\tilde{\rho}$ is a traceless and normalized operator according to the Hilbert-Schmidt norm, i.e., $\tr(\tilde{\rho})=0$ and $\tr(\tilde{\rho}^2)=1$. A simple calculation shows that $r$ can be written as function of the purity $R(\rho)$ [Eq.~\eqref{purity}] as
\begin{equation}\label{rPurity}
r=\sqrt{R(\rho)-R(\rho_0)}.,
\end{equation}
where $R(\rho_0)=1/(2j+1)$ is the purity of the MMS. The maximum value $\sqrt{2j/(2j+1)}$ for $r$ is reached for any pure state, whereas its minimum value $0$ is only reached for the MMS. Around the MMS, there are balls of absolutely classical states. A lower bound for the maximal radius of such balls has been obtained in Ref.~\cite{2017Giraud} and reads as
\begin{equation}\label{rmax}
r_{\mathrm{max}}=\frac{\left[(4j+1)\tbinom{4j}{2j}-(j+1)\right]^{-1 / 2}}{\sqrt{4j+2}}.
\end{equation}
The lower bound \eqref{rmax} decreases exponentially for large spin quantum numbers as $\sqrt[4]{\pi}(2j)^{-\frac{3}{4}}\,2^{-2j-1}$. When $r\leqslant r_{\mathrm{max}}$, the state \eqref{definition_r} is guaranteed to be absolutely classical.

\subsection{Anticoherent spin states}
\label{subsecAC}

A spin state $\rho$ will be said to be \textit{anticoherent} if it verifies~\cite{2006Zimba}
\begin{equation}\label{AC1property}
\langle J_{n}\rangle\equiv\mathrm{Tr}[\rho (\mathbf{J}\boldsymbol{\cdot}\mathbf{n})]=0
\end{equation}
for any unit vector $\mathbf{n}\in\mathbb{R}^3$. From a physical point of view, anticoherent states are the most isotropic spin states, since the spin expectation value $\langle \mathbf{J}\rangle=0$ does not identify a preferred direction in space. The defining property \eqref{AC1property} is that of anticoherence to order 1. It can be generalized to higher orders of anticoherence by requiring that higher moments of $(\mathbf{J}\boldsymbol{\cdot}\mathbf{n})$ are independent of $\mathbf{n}$, thereby reinforcing the isotropic character of the state. For example, anticoherent states to order 2 are defined as spin states for which $\mathrm{Tr}[\rho (\mathbf{J}\boldsymbol{\cdot}\mathbf{n})]=0$ and $\mathrm{Tr}[\rho (\mathbf{J}\boldsymbol{\cdot}\mathbf{n})^2]=c$ with $c>0$ independent of $\mathbf{n}$. More generally, for any integer $q>0$, a state $\rho$ is anticoherent to order $q$, or simply $q$ anticoherent, if $\mathrm{Tr}[\rho (\mathbf{J}\boldsymbol{\cdot}\mathbf{n})^q]$ does not depend on $\mathbf{n}$. Paradigmatic examples of spin-anticoherent states to order 1 (but not higher) are Schrödinger cat states $\frac{1}{\sqrt{2}}(|j,j\rangle +|j,-j\rangle)$ and the states $|j,0\rangle$ for integer $j$.

Anticoherent states are best described in the multipole operator basis, where a state is anticoherent to order $q$ if and only if its state multipoles vanish for $L=1,2,\ldots,q$~\cite{Gir15}, or
\begin{equation}\label{defACq}
\begin{array}{c}
\rho \mathrm{~is~}q\mathrm{-anticoherent}\\[2pt]
\Leftrightarrow\\[2pt]
\rho_{LM}=0\quad \parbox[t]{3cm}{$\forall\; L=1,\ldots,q,$\\ $\forall\;M=-L,\ldots, L$.}
\end{array}
\end{equation}
A $q$-anticoherent state has therefore a multipolar expansion of the form
\begin{equation}
\rho=\rho_0+\sum_{L=q}^{2j}\sum_{M=-L}^{L}\rho_{LM}\,T_{LM}.
\end{equation}
According to our definition, anticoherent spin states can be pure or mixed. While mixed states can be anticoherent to any order (think of the maximally mixed state), pure states can only have a limited order of anticoherence. Numerical searches have shown that the highest achievable order of anticoherence $q_\mathrm{max}$ for pure spin-$j$ states scales approximately as $q_\mathrm{max}\sim \sqrt{4j}$~\cite{Bag15,ACstates}. In this work, we use the acronym \HOAP to refer to pure states with the highest achievable order of anticoherence (highest-order anticoherent pure states). A few examples of \HOAP states are given in Table~\ref{tabACstates} at the end of the paper, just before the Appendixes. Note that \HOAP states are generally not unique for a given spin quantum number, even up to a rotation.

\setlength{\tabcolsep}{5pt}
\renewcommand{\arraystretch}{1.5}

\begin{widetext}
\begin{center}
\begin{table}
\begin{centering}
\begin{tabular}{|c||c|}
\hline 
Single spin-$j$ & Multiple spin-$1/2$ or qubits \tabularnewline
\hline 
\hline
spin quantum number $j=\frac{N}{2}$ & number of qubits $N=2j$
\tabularnewline
\hline 
spin operators & \parbox[t]{7cm}{\centering collective spin operators\\ (irrep with maximal angular momentum $j$)\\[5pt]}
\tabularnewline
\hline 
\parbox[t]{3.5cm}{\centering standard basis $\{|j,m\rangle\}$\\[2pt] with $m=-j,\ldots,j$\\[5pt] }& \parbox[t]{7cm}{\centering symmetric Dicke basis $\{|D_N^{(k)}\rangle\}$\\[2pt] with $k=0,\ldots,N$\\[5pt] }
\tabularnewline
\hline 
\parbox[t]{4.5cm}{\centering magnetic quantum number\\[2pt]  $m=\frac{N}{2}-k$} & \parbox[t]{4.5cm}{\centering number of excitations\\[2pt]  $k=j-m$\\[5pt] }  
\tabularnewline
\hline 
full Hilbert space & symmetric subspace
\tabularnewline
\hline 
(absolutely) classical & (absolutely) separable
\tabularnewline
\hline 
non-classical & entangled
\tabularnewline
\hline 
spin coherent state, $\ket{\n}$ & pure symmetric separable state, $|\Omega\rangle^{\otimes N} $
\tabularnewline
\hline 
\parbox[t]{7cm}{\centering \vskip0pt pure anticoherent\\ state to order $q$} & \parbox[t]{7cm}{\centering pure maximally entangled symmetric state\\ (state with maximally mixed $q$-qubit\\ reductions in the symmetric sector)\\[5pt]}
\tabularnewline
\hline 
\parbox[t]{7cm}{\centering purity-based measure of\\ anticoherence to order $q$, $\mathcal{A}_q$\\[5pt]} & \parbox[t]{7cm}{\centering rescaled linear entropy of the\\ $q$-qubit reduced density matrix\\[5pt]} 
\tabularnewline
\hline 
rotation & symmetric local unitary transformation
\tabularnewline
\hline 
\end{tabular}
\caption{Dictionary of correspondence between spin-$j$ states and symmetric states of $N$ spin $1/2$ or qubits. The constituent spin $1/2$ can be real or fictitious.
\label{tabmapping}}
\par\end{centering}
\end{table}
\end{center}
\end{widetext}

\subsection{Symmetric multiqubit states}\label{SymmMultiQubit}

The Hilbert space associated to an individual spin-$j$ quantum system, $\mathcal{H}\simeq \mathbb{C}^{2j+1}$, is isomorphic to the symmetric subspace of the Hilbert space associated to a system of $N=2j$ spin $1/2$ or qubits. This is due to the fact that $N$ spin $1/2$ can always be coupled to form a collective spin $j=N/2$. The states resulting from this coupling can be shown to be necessarily invariant under permutation of the spins, i.e., symmetric. Therefore, all concepts pertaining to spin states can be reformulated in terms of multiqubit symmetric states (see Table \ref{tabmapping}).

In this framework, standard angular momentum basis states $\ket{j,m}$ can be viewed as symmetric Dicke states of $N=2j$ qubits with $k=j-m$ excitations (number of qubits in the state $\ket{1}$)
\begin{equation}
\label{Dicke}
\ket{D_N^{(k)}} = \frac{1}{\sqrt{\tbinom{N}{k}}} \sum_{\pi} \ket{\underbrace{0\ldots 0}_{N-k}\underbrace{1\ldots 1}_{k}},
\end{equation}
where the sum runs over all distinct permutations $\pi$ of the qubits. Spin-coherent states $\ket{\n}$ can be viewed as $N$-qubit symmetric separable pure states, which are necessarily of the form $\ket{\Omega}^{\otimes N}$ with $\ket{\Omega}$ some single-qubit state. In contrast, it was shown in~\cite{Gir15} that anticoherent spin states are characterized by the fact that their $q$-qubit reduced density matrices are the maximally mixed state in the symmetric sector, $\rho_0 = \mathbb{1}/(q+1)$. As a result, it is possible to quantify the degree of $q$ anticoherence of a pure spin-$j$ state $|\psi\rangle$ using a measure of anticoherence based on purity, which can be related to the Hilbert-Schmidt distance of the reductions of $|\psi\rangle$ to the MMS $\rho_0$~\cite{Bag17}. This measure of anticoherence $\mathcal{A}_{q}(|\psi\rangle)$ is defined by
\begin{equation}
\mathcal{A}_{q}(|\psi\rangle)=\frac{q+1}{q}\left[1-R(\rho_{q})\right]\,,\label{ACR}
\end{equation}
where $R(\rho_{q})$ is the purity of $\rho_{q}=\tr_{\neg q}\left[|\psi\rangle\langle\psi|\right]$, the $q$-qubit reduced density matrix of the state $|\psi\rangle$ viewed as an $N$-qubit symmetric state. The measures of anticoherence $\mathcal{A}_{q}(|\psi\rangle)$ for $q=1,2,\ldots$ are merely the linear entropy of entanglement of the state $|\psi\rangle$ with respect to the possible bipartition $(q,N-q)$ of the $N$ qubits. These measures are obviously rotationally invariant. Spin-coherent states are the only pure states characterized by pure reduced states and are therefore the only states for which $\mathcal{A}_{q}=0$. In contrast, $q$-anticoherent states are characterized by $\rho_q=\rho_0$ and are therefore the only states such that $\mathcal{A}_{q}=1$. For all other states, $\mathcal{A}_{q}$ is strictly between $0$ and $1$~\cite{Bag17}. Let us illustrate these measures on states that will be of interest to us later and which are listed in Table~\ref{tabstates}. GHZ states for any number of qubits and balanced Dicke states for an even number of qubits are both anticoherent to order $1$, but not to order $2$, because we have
\begin{equation*}
\mathcal{A}_{1}(|\mathrm{GHZ}\rangle)=1,\quad \mathcal{A}_{2}(|\mathrm{GHZ}\rangle)=\frac{3}{4}<1,
\end{equation*}
and 
\begin{equation*}
\mathcal{A}_{1}(\ket{\mathrm{DB}})=1,\quad \mathcal{A}_{2}(\ket{\mathrm{DB}})=\tfrac{3(\frac{N}{2}-1)(5\frac{N}{2}-1)}{4 (N-1)^2}<1.
\end{equation*}
In contrast, for $N>2$, the W state is never $1$-anticoherent, whereas the \HOAP states always are since
\begin{equation*}
\mathcal{A}_{1}(|\mathrm{W}\rangle)=2\,\tfrac{N-1}{N^2}<1,\quad \mathcal{A}_{1}(|\mathrm{\HOAP}\rangle)=1.
\end{equation*}
Then, the notion of non-classicality of spin states translates into the notion of multiqubit entanglement for symmetric states. Indeed, a classical state $\rho$ that can be written as \eqref{rhoPfuncgen} with a non-negative $P_{\rho}$ function everywhere is a continuous convex mixture of symmetric separable states $(\ket{\Omega}\bra{\Omega})^{\otimes N}$. Hence, it is separable~\cite{2005Korbicz}. Likewise, absolute classicality of a spin state translates into absolute separability of multiqubit symmetric states~\cite{Zyc01}.

\renewcommand{\arraystretch}{1.6}
\begin{table}[hbt!]
\begin{centering}
\begin{tabular}{|c|c|}
\hline 
Abbreviation & State \\ 
\hline  \hline 
GHZ & \parbox{4.5cm}{\vskip 2pt $|\mathrm{GHZ}\rangle=\frac{1}{\sqrt{2}}(|D_N^{(0)}\rangle +|D_N^{(N)}\rangle)$ \vskip 2pt } \\ 
\hline 
DB & $|\mathrm{DB}\rangle=|D_N^{(\lfloor N/2\rfloor)}\rangle$ \\ 
\hline 
W & $|\mathrm{W}\rangle=|D_N^{(1)}\rangle$ \\ 
\hline 
\HOAP & \parbox{4.4cm}{\vskip 4pt Highest-Order Anticoherent Pure states, see Table~\ref{tabACstates} and Refs.~\cite{2006Zimba,Bag15,Bjo15,ACstates}\vskip 4pt} \\
\hline
\end{tabular} 
\caption{Families of $N$-qubit symmetric states of interest for this work, expressed in terms of Dicke states \eqref{Dicke}. In the definition of the balanced Dicke state $\ket{\mathrm{DB}}$, $\lfloor N/2\rfloor$ denotes the largest integer smaller than or equal to $N/2$.\label{tabstates}}
\par\end{centering}
\end{table}

\section{depolarization dynamics}\label{Sec:CollectiveDecoherence}

We consider a spin-$j$ system undergoing (Markovian) depolarization along the three spatial directions $x$, $y$, and $z$ with \emph{a priori} different rates $\gamma_x$, $\gamma_y$, and $\gamma_z$ (see Eq.~\eqref{MEq}). In Appendix \ref{Appendix_models}, we give examples of physical models that provide a microscopic basis for our master equation: a spin interacting with a fluctuating magnetic field and an ensemble of two-level atoms interacting with the electromagnetic field at infinite temperature. Quantum light states can also undergo depolarization due to random SU($2$) rotations as light propagates through an optical fiber whose birefringent index fluctuates with time~\cite{2014Rozema}. Many more models leading to the same form of dynamical evolution have been proposed and studied in the literature. The derivation of a master equation describing isotropic depolarization of multiphoton states was carried out in~\cite{2013Rivas} based on SU($2$) invariance. The effect of isotropic decoherence for a qubit on Uhlmann's mixed-state geometric phase was studied in~\cite{Tidstr2003}. In Ref.~\cite{2006Klimov}, the anisotropic depolarization of light stemming from its interaction with a material medium was analysed. A master equation describing anisotropic depolarization of a spin $1/2$ was derived from a microscopic Hamiltonian model in~\cite{2017Arsenijevic} and from disordered Hamiltonian ensembles in~\cite{2021Chen}. A non-Markovian version for a central spin $1/2$ interacting with a spin bath has been derived in~\cite{2017Bhattacharya}.
Finally, let us note that a weak continuous measurement of the Cartesian components of a spin leads to the same form of dynamical evolution and offers the possibility of tuning the three depolarization rates $\gamma_x$, $\gamma_y$ and $\gamma_z$. On the other hand, microscopic models and basic thermodynamic principles can lead to constraints on the possible values that rates can take, as we explain in the next subsection.

In Sec.~\ref{sec:Master_eq}, we present the master equation (\ref{MEq}) central to this work. In Sec.~\ref{sec:METPB}, we write the master equation in the \basis. We then discuss a conservation law for the dynamics (Sec.~\ref{sec:ACConservation}), its steady states (Sec.~\ref{sec:statstate}), and provide a general solution for the density matrix, its reductions, and its purity in Sec.~\ref{Sec:solME} and~\ref{Sec:purityDecrease}. We emphasize that particular instances of these solutions were already known (see, e.g., ~Refs.~\cite{2009Muller,2013Rivas,BreuerPetruccione}).

\subsection{Master equation}\label{sec:Master_eq}
We consider the dynamics of a spin governed by a master equation of the Lindblad form
\begin{equation}\label{MEq}
\dot{\rho}(t)=\frac{i}{\hbar}\left[\rho(t),H\right]+\sum_{\alpha=x,y,z}\mathcal{D}_{\alpha}[\rho(t)]
\end{equation}
with $H=\hbar\omega J_{z}$ and
\begin{equation}\label{Dissalpha}
\begin{aligned}
\mathcal{D}_{\alpha}[\rho] &= \gamma_{\alpha}\big(2J_{\alpha}\rho J_{\alpha}-J_{\alpha}^2\rho-\rho J_{\alpha}^2\big)
\end{aligned}
\end{equation}
where $\mathbf{J}=(J_x,J_y,J_z)$ is the spin operator in the spin-$j$ sector. In particular, our master equation describes the well-known pure dephasing of a spin when $\gamma_x=\gamma_y=0$ and $\gamma_z\ne0$. In the following, we will refer to \emph{isotropic} depolarization when $\gamma_x=\gamma_y=\gamma_z$ and to \emph{anisotropic} depolarization when $\gamma_x=\gamma_y\ne\gamma_z$. Since the dissipator contains only Hermitian jump operators, the Lindblad generator is unital and it follows that the purity of the state $\rho(t)$ can only decrease with time (see, e.g., \cite{Lid05}). As highlighted earlier in Refs.~\cite{2009Muller,2013Rivas}, isotropic depolarization as described by Lindblad's master equation \eqref{MEq} has richer dynamics than when described by the standard phenomenological SU($N+1$) depolarization channel
\begin{equation}\label{depolchannel}
\mathcal{E}(\rho)=(1-p)\, \rho+p\, \frac{\mathbb{1}}{N+1}
\end{equation}
with $0 < p \leqslant 1$.

We now come back to the possible constraints on the values that rates can take when obtained from microscopic models. Under the assumptions that the spin environment is in thermal equilibrium and that the interaction Hamiltonian between the spin and the environment commutes with both Hamiltonians of the spin and the environment, a condition coined as strict conservation of energy, the unitary and dissipative evolutions necessarily commute~\cite{2013Kosloff, 2020Kosloff}, i.e., $\left[\mathcal{U},\mathcal{D}\right] = 0$ where $\mathcal{U}(\boldsymbol{\cdot})=(i/\hbar)\left[\boldsymbol{\cdot},H\right]$ and $\mathcal{D}(\boldsymbol{\cdot})=\sum_{\alpha=x,y,z} \mathcal{D}_\alpha(\boldsymbol{\cdot})$  for the master equation \eqref{MEq}. For $\omega\ne 0$, the condition $\left[\mathcal{U},\mathcal{D}\right] = 0$ is fulfilled only when $\gamma_x=\gamma_y$. A direct calculation indeed shows that
\begin{equation}
\left[\mathcal{U}(\rho),\mathcal{D}(\rho)\right] = \tfrac{i\omega}{2}(\gamma_x-\gamma_y)\sum_{LM}\Big(\sum_\pm \pm d_{LM}^\pm T_{LM\pm 2}\Big)\rho_{LM}
\end{equation}
with $d_{LM}^\pm$ defined in Eq.~\eqref{dpm} and $\rho_{LM}$ the state multipoles of $\rho$. Hence, the conclusion that $\left[\mathcal{U}(\rho),\mathcal{D}(\rho)\right] = 0$ for any $\rho$ and $\omega\ne 0$ only if $\gamma_x=\gamma_y$. Physically, this can be understood by observing that the unitary evolution induces a precession of the spin around the $z$-axis and that the dissipative evolution is invariant under a rotation around the $z$-axis only when $\gamma_x=\gamma_y$. For this reason, we will concentrate from Sec.~\ref{sec:statstate} on the case $\gamma_x=\gamma_y\equiv \gamma_{\perp z}$. For isotropic depolarization, we have that $\mathcal{U}$ commutes with $\mathcal{D}$ even in the more general case of a Hamiltonian of the form $H=\hbar\omega_x J_x+\hbar\omega_y J_y+\hbar\omega_z J_z$. When Hamiltonian evolution and dissipative evolution commute, the former has no impact on how the purity and entanglement of a state evolve with time, because the unitary evolution with a linear Hamiltonian in the spin operators itself induces a rotation (or local symmetric unitary transformation) which preserves the purity and entanglement.

In the rest of this work, we will adopt the view of a spin state as a multiqubit symmetric state, so that $J_{\alpha}=\tfrac{1}{2}\sum_{i=1}^{N}\sigma_{\alpha}^{(i)}$ are now collective spin operators with $\sigma_{\alpha}^{(i)}$ the Pauli operators $\sigma_{\alpha}$ for qubit $i$, and the dynamics corresponds to the collective decoherence of $N=2j$ qubits, initially in a pure symmetric state. The master equation (\ref{MEq}) preserves the permutation symmetry of the initial state, so that we are in fact dealing with an irreducible representation of dimension $N+1$ of the collective spin operators.

\subsection{Master equation in the multipole operator basis}\label{sec:METPB}

The system's state can be expanded at all times in the multipole operator basis $\{T_{LM}\}$ (\basis) as
\begin{equation}\label{rhoexpTLM}
\rho(t)=\sum_{L=0}^{N}\sum_{M=-L}^{L}\rho_{LM}(t)\,T_{LM},
\end{equation}
where the state multipoles $\rho_{LM}$ evolve according to (see Appendix \ref{Appendix_MEMOB})
\begin{equation}\label{eq:ME_TLM}
\begin{aligned}
\dot{\rho}_{LM} = & -\left(\Gamma_{LM}+i\,\omega M\right)\,\rho_{LM} \\
& -\Gamma_{LM+2}\,\rho_{LM+2}-\Gamma_{LM-2}\,\rho_{LM-2},
\end{aligned}
\end{equation}
where
\begin{equation}\label{offdiagrate}
\begin{aligned}
& \Gamma_{LM}=\gamma_{z}\,M^{2}+\frac{\gamma_{x}+\gamma_{y}}{2}\left[ L(L+1)-M^{2}\right],\\
& \Gamma_{LM\pm 2}=\frac{\gamma_{x}-\gamma_{y}}{4}\,d_{LM}^{\pm}
\end{aligned}
\end{equation}
with
\begin{equation}\label{dpm}
\begin{aligned}
d_{LM}^{\pm} ={}& \sqrt{(L\mp M)(L\pm M+1)}\\
& \times \sqrt{(L\mp M-1)(L\pm M+2)}.
\end{aligned}
\end{equation}
We note that state multipoles with different values of $L$ are decoupled, as well as matrix elements with even and odd values of $M$. This means that the Liouvillian superoperator is represented in the \basis~by a block-diagonal matrix, with $N+1$ blocks of size $2L+1$ for $L=0,\ldots,N$. From Eq.~\eqref{eq:ME_TLM}, we see that the block for $L=0$ is equal to $0$, so that the stationary state has always a component on the maximally mixed state in the symmetric sector, $\rho_0=\mathbb{1}/(N+1)$. This reflects the fact that the component $\rho_{00}$ on $T_{00}=\mathbb{1}/\sqrt{N+1}$ remains equal to $1/\sqrt{N+1}$ at all times, which is simply due to normalization of the state $\rho$. The blocks for $L>0$ have elements $-\Gamma_{LM}$ for $M=-L,\ldots,L$ on the main diagonal, $-\Gamma_{LM+2}$ along the $2$-diagonal, and $-\Gamma_{LM-2}$ along the $-2$-diagonal. Because of the relation $d_{LM}^{+} =d_{LM+2}^{-}$ [see Eq.~\eqref{dpm}], the $2$-diagonal and the $-2$-diagonal are equal. When $\gamma_{x}=\gamma_{y}$, $\Gamma_{LM\pm 2}=0$, the Liouvillian is diagonal in \basis and each $\rho_{LM}$ evolves independently.

\subsection{Conservation of anticoherence}\label{sec:ACConservation}

The dynamics generated by the master equation \eqref{eq:ME_TLM} conserves the property of anticoherence of a state at any time. Indeed, if the state multipoles $\rho_{LM}$ are initially equal to zero for some $L$ and $\forall\;M=-L,\ldots, L$, they remain zero at all times because there is no coupling between blocks corresponding to different values of $L$. Therefore, an initial $q$-anticoherent state characterized by the property \eqref{defACq} remains $q$ anticoherent at any time. We emphasize that this property would still hold if the $\gamma_{\alpha}$ rates were time dependent. Therefore, the property of anticoherence of a state is very generally conserved under the model of decoherence corresponding to the master equation \eqref{MEq}.

\subsection{Stationary state}\label{sec:statstate}

When $\gamma_{x}=\gamma_{y}\equiv \gamma_{\perp z}=0$ and $\gamma_{z}\ne 0$, Eq.~\eqref{eq:ME_TLM} shows that the components $\rho_{L0}$ are conserved for all $L$ while all other components decay to $0$. Starting from an initial state $\rho(0)$, the stationary state is then given by
\begin{equation}
\begin{aligned}
\rho(\infty) =\sum_{L=0}^{N}\rho_{L0}(0)\,T_{L0}= \sum_{m=-j}^{j} \rho_{mm}(0)\left|j, m\right\rangle\left\langle j, m\right|
\end{aligned}
\end{equation}
with $\rho_{mm}(0)\equiv\left\langle j, m\right|\rho(0) \left|j, m\right\rangle$ because $T_{L0}=\sqrt{\frac{2 L+1}{2j+1}} \sum_{m} \CG{j}{m}{L}{0}{j}{m} \left|j, m\right\rangle\left\langle j, m\right|$ where $C_{m_1m_2m}^{j_1j_2j}\equiv \langle j_1 m_1 j_2 m_2 |j,m\rangle$ are Clebsch-Gordan coefficients following the notation of~\cite{Louck}. The set of projectors $\{\left|j, m\right\rangle\left\langle j, m\right|\}$ is thus the pointer basis in this case.

When $\gamma_{\perp z}\ne 0$, Eq.~\eqref{eq:ME_TLM} shows that each component $\rho_{LM}$ evolves separately and decays to $0$ (except for $\rho_{00}$) at a rate $\Gamma_{LM}>0$. The only stationary state is then the MMS [Eq.~\eqref{MMS}].

\subsection{Exact solutions for $\rho$ and its reductions}
\label{Sec:solME}

Under the condition of strict conservation of energy, $\gamma_{x}=\gamma_{y}$ and Eq.~\eqref{offdiagrate} yields $\Gamma^{\pm}_{LM\pm 2}=0$, which implies that the state multipoles $\rho_{LM}$ are all decoupled from each other. They thus evolve according to 
\begin{equation}\label{equrhoLMdecoup}
\dot{\rho}_{LM}=-\left(\Gamma_{LM}+i\,\omega M\right)\,\rho_{LM}
\end{equation}
whose general solution is given by
\begin{equation}\label{gensoldiag}
\rho_{LM}(t)=e^{-\left(\Gamma_{LM}+i\,\omega M\right) t}\rho_{LM}(0).
\end{equation}
All state multipoles have their own diﬀerent decay rates, which contrasts with the phenomenological description of depolarization through the SU($N+1$) depolarization channel \eqref{depolchannel} for which all state multipoles with $L\neq 0$ decrease with the same factor.
The solution of the master equation in the Dicke basis readily follows by expressing the matrix elements $\langle j,m|\rho|j,m'\rangle$ in terms of the state multipoles $\rho_{LM}$ and using Eq.~\eqref{gensoldiag}. We get~\cite{Varshalovich}
\begin{equation}\label{gensoldiagjm}
\langle j,m|\rho|j,m'\rangle=\sum_{L=0}^{2j}\sqrt{\tfrac{2L+1}{2j+1}}\CG{j}{j-m'}{L}{m'-m}{j}{j-m}\,\rho_{Lm'-m}
\end{equation}
Alternatively, using Eq.~\eqref{PfunctrhoLM}, the state $\rho(t)$ can be expressed in terms of the time-dependent $P$ function
\begin{equation}\label{Prhot}
P_{\rho(t)}(\Omega)=\sum_{L,M} \alpha_{L}^{(N)} \rho_{LM}(0)\,e^{-\left(\Gamma_{LM}+i\,\omega M\right) t} \,Y_{LM}(\Omega).
\end{equation}

The solution \eqref{gensoldiag} allows us to directly access the state multipoles of all the reduced states of $\rho$. In the Appendix \ref{Appendix_MOBptrace}, we show that the state multipoles $\rho_{LM}^{(q)}$ of the $q$-qubit reduced density matrix $\rho_q=\tr_{\neg q}\left[\rho\right]$ can be expressed in terms of the $\rho_{LM}$ of $\rho$ as 
\begin{equation}\label{statemultipolerhoq}
\rho_{LM}^{(q)}(t)=\frac{q!}{N!}\sqrt{\tfrac{(N-L)!(N+L+1)!}{(q-L)!(q+L+1)!} }\,\rho_{LM}(t)
\end{equation}
for $L\leqslant q$ and $-L\leqslant M\leqslant L$.

Note that in the case where depolarization rates are time dependent and possibly negative, in which case Eq.~\eqref{MEq} must be considered as a time-convolutionless (TCL) non-Markovian master equation, the previous developments are still valid, provided that $\Gamma_{LM}\, t$ in the solution \eqref{gensoldiag} is replaced by $\int_0^t\Gamma_{LM}(t')\, dt'$. When rates are negative, the Liouvillian is no longer unital and the purity does not necessarily decrease monotonically over time. In what follows, Eqs.~\eqref{gensoldiag} and \eqref{statemultipolerhoq} are central to our analytical developments and to all our numerical results.

\subsection{Decrease of purity}\label{Sec:purityDecrease}
In order to identify the states that are the most prone to decoherence, we focus on the rate of change of the purity $\dpuritydt(\rho)$ and its higher-order time derivatives. These quantities can be expressed in a simple form when $\gamma_{x}=\gamma_{y}$ because then Eq.~\eqref{equrhoLMdecoup} implies that $d|\rho_{LM}|^2/dt=-2\left|\rho_{LM}\right|^{2}\Gamma_{LM}$. Repeated use of the latter equality together with Eq.~\eqref{purityrhoLM} yields for the $n$th-order time derivative
\begin{equation}
\frac{d^n\purity(\rho)}{dt^n}=(-2)^n\sum_{L,M}|\rho_{LM}|^2\,(\Gamma_{LM})^n.
\end{equation}
This equation can be rewritten as
\begin{equation}
\frac{d^n\purity(\rho)}{dt^n}=(-2)^n\,\overline{(\Gamma_{LM})^n}\,\purity(\rho),
\end{equation}
where we define the average value in state $\rho$ of a function $f$ of variables $L$ and $M$ by
\begin{equation}\label{averageML}
\overline{f(L,M)}=\frac{\sum_{L,M}\left|\rho_{LM}\right|^{2}f(L,M)}{\sum_{L,M}\left|\rho_{LM}\right|^{2}}.
\end{equation}

In particular, the rate of change of purity, which is also equal to the opposite of the linear entropy production rate~\cite{Hor09}, is given by
\begin{equation}\label{Rdott0}
\begin{aligned}
\dpuritydt(\rho) =  -2\,\overline{\Gamma_{LM}} \,\purity(\rho) \leqslant 0.
\end{aligned}
\end{equation}
Similarly, the second time derivative of $\purity(\rho)$ is given by
\begin{equation}\label{Rddott0}
\begin{aligned}
\ddot{\purity}(\rho)=4\,\overline{(\Gamma_{LM})^2} \,\purity(\rho)\geqslant 0.
\end{aligned}
\end{equation}
For the maximally mixed state, the purity does not change over time and we have $\frac{d^n\purity(\rho)}{dt^n}=0$ for all $n$.

More generally, the solution \eqref{gensoldiag} for the state multipole of $\rho(t)$ along with Eq.~\eqref{purityrhoLM} for the purity gives
\begin{equation}
R(t)=\frac{1}{N+1}+\sum_{L>0,M} |\rho_{LM}(0)|^2\; e^{-2\,\Gamma_{LM}\,t},
\label{puritytanisotropic}
\end{equation}
where we have set $R(t)\equiv R(\rho(t))$. All the terms in Eq.~\eqref{puritytanisotropic} are positive. Therefore, for any integer $q>0$, the purity can be lower bounded by retaining only the terms of the sum with $L>q$, that is
\begin{align}
R(t) \geqslant \frac{1}{N+1}+\sum_{L>q,M} |\rho_{LM}(0)|^2\; e^{-2\,\Gamma_{LM}\,t}.\label{puritytisotropic}
\end{align}
For pure states, a state-independent bound can be obtained by lower bounding each decreasing exponential in Eq.~\eqref{puritytanisotropic} by the fastest decreasing one and using the purity condition $\sum_{L,M}\left|\rho_{LM}\right|^{2}=1$. This yields
\begin{equation}
R(t)\geqslant\frac{1}{N+1}+\frac{N}{N+1}\; e^{-2\,\Gamma_{LM}^{\max}\,t}.
\label{purityboundstateindependent}
\end{equation}
with $\displaystyle\Gamma_{LM}^{\max}=\max\{\Gamma_{LM}\}_{L,M}$.

\section{Isotropic depolarization}
\label{Sec:extrisotrop}

In this section, we focus on isotropic depolarization defined by equal rates in the three spatial directions $\gamma_{x}=\gamma_{y}=\gamma_{z}\equiv\gamma$. This model is commonly used as a phenomenological model to describe decoherence in quantum computers when errors arise from uncontrolled small random SU($2$) rotations~\cite{Nielsen_Chuang}. In Sec.~\ref{sec:puritiesIsotropic} and~\ref{sec:Superdecoherence}, we study the evolution of purities and clarify the condition for the occurence of superdecoherence, respectively. In Sec.~\ref{sec:extremalStatesIsotropic}, we identify the states most susceptible to decoherence. Sec.~\ref{sec:QSL} is concerned with quantum speed limits and in Sec.~\ref{sec:entanglementIsotropic}, we study how entanglement is gradually lost over time due to depolarization.

\subsection{Evolution of purities}\label{sec:puritiesIsotropic}

We start with the rate of change of the purity of the global state $\rho$, which is given by
\begin{equation}\label{Rdotisotropic}
\dpuritydt(\rho)  = -2\,\overline{ L(L+1)}\,\gamma\,\purity(\rho)\leqslant 0
\end{equation}
and can be rewritten as (see Appendix \ref{Appendix_reducedpurity})
\begin{equation}\label{rateRrho}
\dpuritydt(\rho) = -2\gamma\left[N(N+1)\purity(\rho) - N^2\purity(\rho_{N-1}),\right]
\end{equation}
where $\purity(\rho_{N-1})$ is the purity of the $(N-1)$-qubit reduced density matrix of $\rho$. Due to the proportionality of the state multipoles expressed by the relation~\eqref{statemultipolerhoq} and the diagonal form of the master equation in the \basis, the master equation keeps exactly the same form \eqref{equrhoLMdecoup} for the reduced states $\rho_q$ than for the global state $\rho$. Using \eqref{rateRrho}, we can thus write the closed set of equations for the purities
\begin{equation}\label{setequR}			
\begin{aligned}
& \dpuritydt(\rho_{1}) = -2\gamma\left[2\purity(\rho_{1}) - 1\right], \\
& \dpuritydt(\rho_{2}) = -2\gamma\left[6\purity(\rho_{2}) - 4\purity(\rho_{1})\right],\\
& \qquad\vdots \\
& \dpuritydt(\rho) = -2\gamma\left[N(N+1)\purity(\rho) - N^2\purity(\rho_{N-1})\right].
\end{aligned}
\end{equation}
The first $q$ equations of the system \eqref{setequR} also form a closed set of equations. Therefore, the time evolution of $\purity(\rho_q)$ depends only on the initial purities of the reduced states of at most $q$ qubits, regardless the total number of qubits $N$. Analytical solutions for $\purity(\rho_q)$ are given for the smallest values of $q$ in Appendix~\ref{Appendix_purities}.

\subsection{Superdecoherence}\label{sec:Superdecoherence}

It is known that collective dephasing can lead to a phenomenon called superdecoherence~\cite{2019Kattem, 2005Berman, 2011Monz}, where the rate at which states lose their coherence scales as $N^{2}$ with $N$ the number of qubits. In this subsection, we establish the conditions for superdecoherence to occur when the qubits are subject to collective isotropic depolarization.

Let us set $R(t)\equiv R(\rho(t))$ and denote by $T$ the time it takes for the purity of a state to decrease to half its initial value $R(0)$. A lower bound $T_\mathrm{min}$ on $T$ can be obtained by replacing $R(\rho_{N-1})$ in Eq.~\eqref{rateRrho} by its minimum value $1/N$, solving the resulting differential equation for $R(t)$ and then solving $R(T_\mathrm{min})=R(0)/2$ for $T_\mathrm{min}$. This yields the time
\begin{equation}\label{boundT}
T_\mathrm{min}=\frac{\ln \left(2+\frac{2}{(N+1) R(0)-2}\right)}{2 \gamma  N (N+1)}\widesim{N\gg 1} \frac{\ln 2}{2 \gamma N^2}
\end{equation}
such that $R(t)>R(0)/2$ for $t<T_\mathrm{min}$. The lower bound $T_\mathrm{min}$ scales as $N^{-2}$, which leaves room for superdecoherence.
In fact, exact conditions for the occurrence of superdecoherence can be found by rewriting Eq.~\eqref{rateRrho} as
\begin{equation}\label{ratedecoh}
\dpuritydt(\rho) =-2\gamma \purity(\rho)  N - 2\gamma \left[\purity(\rho)-\purity(\rho_{N-1})\right] N^2
\end{equation}
where the linear and quadratic terms in $N$ have been separated. We now show that the coefficient in front of $N^2$ can only be negative for entangled states, a result that can be interpreted as saying that \emph{isotropic superdecoherence cannot occur without entanglement}. To this end, we use the entanglement criterion based on the Rényi entropy $S_q$ for $q=2$ \cite{footnoteRenyi,Zyczkowski_book}, which states that if a state $\rho$ is separable, then $R(\rho_q)\geqslant R(\rho)$ for any number of qubits $q<N$. This criterion reflects the fact that the reduced states of a separable state are always less mixed than the global state is. Therefore, for separable states $\purity(\rho) -\purity(\rho_{N-1})$ is always negative. For example, it is equal to $-1/[N(N+1)]$ for the maximally mixed state. It can be positive only for entangled states. A numerical optimisation shows that $\purity(\rho)-\purity(\rho_{N-1})$ is smaller than or equal to $1/2$, where the value $1/2$ is reached when $\rho$ is a pure $1$-anticoherent state. In this case, the rate of decrease of purity takes its maximal value $\dpuritydt =-\gamma N(N+2)$ scaling quadratically with $N$.  

A consequence of the preceding result is that superdecoherence can never occur when the system starts from a separable state. Indeed, as we show in Appendix \ref{Appendix_coherencepreservation}, isotropic depolarization leads to a separability-preserving (SEPP) dynamical map. This means that separable states can only evolve into separable states, which we have just shown cannot display superdecoherence. Interestingly, the situation is fundamentally different for the closely related phenomenon of superradiance, which can occur without any entanglement~\cite{Wol14}, starting, for example, from the separable state in which all atoms are excited.

For an initial pure state $|\psi_0\rangle$, the rate \eqref{ratedecoh} at $t=0$ can be rewritten as (see Appendix~\ref{Appendix_PureDecoRate} for more detail)
\begin{equation}\label{ratet0}
\begin{aligned}
\dpuritydt\big|_{t=0} & =-2\gamma\left(N+\frac{N^2}{2}\mathcal{A}_1\right)
\end{aligned}
\end{equation} 
where $\mathcal{A}_1$ is the measure of anticoherence to order $1$ of $|\psi_0\rangle$ defined in Eq.~\eqref{ACR} and expressible as
\begin{equation}
\mathcal{A}_1=1 - \frac{1}{j^2}\,|\langle\psi_0|  \mathbf{J} | \psi_0\rangle|^2.
\end{equation} 
Equation~\eqref{ratet0} relates superdecoherence to anticoherence (and thus to entanglement of $|\psi_0\rangle$, see Sec.~\ref{SymmMultiQubit}) because the term scaling quadratically with the number of qubits is directly proportional to $\mathcal{A}_1$. This term vanishes for spin-coherent states ($\mathcal{A}_1=0$) and is maximal for anticoherent states ($\mathcal{A}_1=1$), which consequently display the largest depolarization rate $\dpuritydt\big|_{t=0}=-\gamma N(N+2)$.

Higher-order time derivatives of the purity can be evaluated by repeated use of the set of equations \eqref{setequR}. For example, at any time $t$, we have for the second-order time derivative of $R$
\begin{equation}\label{Rddt0}
\begin{aligned}
\ddot{R} & = -2\gamma\left[N(N+1)\dpuritydt(\rho) - N^2\dpuritydt(\rho_{N-1})\right]\\
& = 4\gamma^2N^2\Big[(N+1)^2\purity(\rho) - 2N^2\purity(\rho_{N-1})\\
& \qquad\qquad\qquad\qquad\quad+(N-1)^2\purity(\rho_{N-2})\Big].
\end{aligned}
\end{equation}
The quantity $\ddot{R}$ does not only depend on $\purity(\rho)$ and $\purity(\rho_{N-1})$ as $\dot{R}$ did, but also on $\purity(\rho_{N-2})$, the purity of the reduced density matrix of $N-2$ qubits. For $\rho$ an initial pure state $\ket{\psi_0}\bra{\psi_0}$, Eq.~\eqref{Rddt0} can be rewritten as
\begin{equation}\label{rate2t0}
\ddot{R}\big|_{t=0} = \frac{4}{3}\gamma^2N^2\left[6+3N^2\mathcal{A}_1-2(N-1)^2\mathcal{A}_2\right],
\end{equation}
where we have set $\mathcal{A}_q=\mathcal{A}_q(|\psi_0\rangle)$ for $q=1,2$. 
It can be seen from Eq.~\eqref{rate2t0} that, among $1$-anticoherent states, $2$-anticoherent states yield the smallest value for $\ddot{R}|_{t=0}$. Equation~\eqref{rate2t0} with $\mathcal{A}_1=\mathcal{A}_2=1$ gives immediately $\ddot{R}|_{t=0}=\frac{4}{3}\gamma^2N^2(N+2)^2$. This means that $2$-anticoherent states not only lead to the highest rate of depolarization $\dot{R}|_{t=0}$, but that this rate also decreases the slowest over time among all such states. In contrast, the maximal value of $\ddot{R}|_{t=0}$ is reached for GHZ states, for which $\mathcal{A}_1=1$, $\mathcal{A}_2=3/4$ and thus $\ddot{R}|_{t=0}=2 \gamma^2 N^2 (N^2+2 N+3)$. This means that among the maximally entangled pure states with respect to the $(1,N-1)$ bipartition, the GHZ states are those for which the depolarization rate decreases most rapidly with time.

\subsection{\HOAP states are extremal} \label{sec:extremalStatesIsotropic}

\begin{figure}[!h]
\begin{centering}
\includegraphics[width=0.475\textwidth]{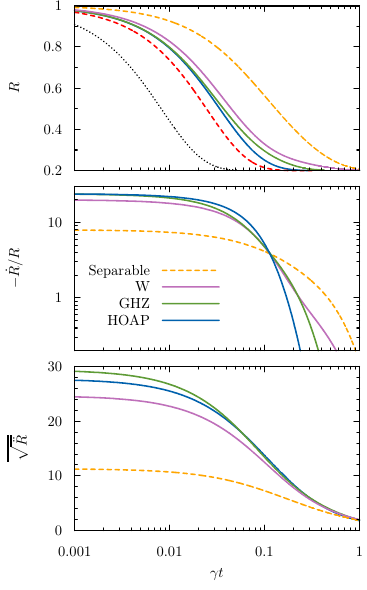}
\par\end{centering}
\caption{Isotropic depolarization for different $4$-qubit states: separable state (orange dashed), W state (purple), GHZ state (green), and \HOAP state given by Eq.~\eqref{eq:anticoherent_state}  (blue). Top: Purity as a function of time, monotonically decreasing towards the asymptotic value $1/5$. The black dotted curve shows the bound \eqref{QSL_Uzdin}, while the red dashed curve shows the improved bound \eqref{purityboundstateindependent}. Middle: Purity loss rate as a function of time. Bottom: Time-average of the square root of the second time derivative of the purity, which is related to the quantum speed limit~\eqref{QSL_Campaioli} giving the bound $\overline{\sqrt{\ddot{R}}} \leqslant  480$ (not shown in the plot). \label{fig:isotropic_R}}
\end{figure}

In the case of isotropic depolarization, the general solution \eqref{gensoldiag} reduces to
\begin{equation}
\rho_{LM}(t)=e^{-\left[\gamma\, L(L+1)+i\,\omega M\right] t}\rho_{LM}(0).
\label{eq:isotropic_decoherence}
\end{equation}
This solution shows that state multipoles with the same angular momentum quantum number $L$ all decay at the same rate $\gamma\, L(L+1)$. Therefore, the initial states that reach the MMS, $\rho_0=\mathbb{1}/(N+1)$, the faster are those whose non-zero state multipoles have high $L$ only. For $q$-anticoherent states [see Eq.~\eqref{defACq}], $\rho_{LM}=0$ $\forall\, L=1,\ldots,q$ and the state multipoles and coherences in the Dicke basis decay over time at a rate at least equal to $\gamma\, (q+1)(q+2)$. Thus, the higher the order of anticoherence $q$ of the initial state, the faster the steady state is reached. This statement can be made more precise by observing that anticoherent states to order $q$ are the only states that saturate the bound \eqref{puritytisotropic}. Therefore, pure anticoherent states to the highest achievable order (\HOAP states introduced in Sec. \ref{subsecAC}) yield, among pure initial states, the lowest purity after \emph{arbitrary evolution time} (see, e.g.,~Fig.~\ref{fig:isotropic_R}). They are extremal for isotropic depolarization.

\emph{At short times}, the pure states most prone to decoherence are, at first order, those for which $\overline{\Gamma_{LM}}$ in Eq.~\eqref{Rdott0} takes the highest possible value at $t=0$. At second order, it is required that $\overline{\Gamma_{LM}}$ is the largest and, at the same time, $\overline{(\Gamma_{LM})^2}$ in Eq.~\eqref{Rddott0} is the smallest at $t=0$. Indeed, the rate $\overline{\Gamma_{LM}}$ calculated at $t=0$ should decrease as slowly as possible over time if one wants the purity loss rate to remain large. As we have shown in the preceding subsection, for isotropic depolarization, $\overline{\Gamma_{LM}}$ is maximal and $\overline{(\Gamma_{LM})^2}$ is minimal for any $q$-anticoherent state with $q\geqslant 2$.

\subsection{Quantum Speed Limits}\label{sec:QSL}

Quantum speed limits (QSLs) reflect constraints on the minimal time required for an initial state to evolve to a target final state under certain dynamics. These limits may also be expressed in terms of quantum speeds. In this work, we consider two QSL for dissipative dynamics recently derived in the literature. The first one gives a bound on the geometric speed, defined as~\cite{2019Campaioli}
\begin{equation}
\overline{\Vert\dot{\rho}\Vert} = \frac{1}{\tau}\int_0^\tau \sqrt{\mathrm{Tr}\left(\dot{\rho}\dot{\rho}^\dagger\right)} \, dt
\end{equation}
where $\tau>0$ is an evolution time, $\Vert \cdot \Vert$ is the Hilbert-Schmidt norm and an overline denotes a time average over the interval $[0,\tau]$. The bound (8) in Ref.~\cite{2019Campaioli} holds for any Lindblad master equation and reduces in our case to
\begin{equation}
\begin{aligned}
	\overline{\Vert\dot{\rho}\Vert}  &{}\leqslant 8 \sum_{\alpha=x,y,z} \gamma_\alpha^2 \Vert J_\alpha\Vert^2\\
	&{} \leqslant \frac{2}{3}\left(\gamma_x^2+\gamma_y^2+\gamma_z^2\right)N(N+1)(N+2)
\end{aligned}
\end{equation}
Moreover, for $\gamma_x=\gamma_y=\gamma_{\perp z}$, $\Vert\dot{\rho}\Vert=\sqrt{\ddot{R}}/2$, so that the QSL reads
\begin{equation}\label{QSL_Campaioli}
	\overline{\sqrt{\ddot{R}}} \leqslant \frac{4}{3}\left(2\gamma_{\perp z}^2+\gamma_z^2\right) N(N+1)(N+2).
\end{equation}
Figure~\ref{fig:isotropic_R} shows $\overline{\sqrt{\ddot{R}}}$ for different $4$-qubit states undergoing isotropic depolarization. At short times, the state with the largest $\overline{\sqrt{\ddot{R}}}$ and thus closest to the upper bound in \eqref{QSL_Campaioli} (equal to $480$) is the GHZ state. This is perfectly consistent with our result \eqref{rate2t0}.

Another QSL particularly relevant to this work is based on the purity of the difference between the state $\rho$ and its stationary state $\rho_0$, the so-called purity deviation $R_D(t)=R(\rho(t)-\rho_0)$~\cite{2016Uzdin}. In the Liouville space, the density operator $\rho$ can be vectorized and its evolution under \eqref{MEq} can be written as $|\dot{\rho}\rangle=\mathcal{H}|\rho\rangle$. Under this evolution, the purity deviation of $\rho$ in a time interval $[t_i,t_f]$ is bounded according to~\cite{2016Uzdin}
\begin{equation}\label{boundUzdin}
	\left| \mathrm{ln}\left(\frac{R_D(t_f)}{R_D(t_i)}\right)\right| \leqslant \int_{t_i}^{t_f} \Vert\mathcal{H}-\mathcal{H}^\dagger\Vert_{\mathrm{sp}} \, dt,
\end{equation}
where $\Vert \cdot\Vert_{\mathrm{sp}}$ denotes the spectral norm~\cite{spectralnorm}. For the master equation \eqref{MEq}, one finds $\Vert\mathcal{H}-\mathcal{H}^\dagger\Vert_{\mathrm{sp}}=\gamma N(N+1)(N+2)$ in the case of isotropic depolarization. By substituting $t_i=0$ and $t_f=t$ into \eqref{boundUzdin} and assuming that the system starts from a pure initial state, the bound can then be expressed in terms of the purity $R(t)$ of $\rho(t)$ as follows
\begin{equation}\label{QSL_Uzdin}
	R(t) \geqslant \frac{1}{N+1}+\frac{N}{N+1}\;e^{-\gamma N(N+1)(N+2)t}.
\end{equation}
The bound in Eq.~\eqref{QSL_Uzdin} is state independent, just like the one in Eq.~\eqref{purityboundstateindependent}. The important difference is that the decay rate in Eq.~\eqref{QSL_Uzdin} scales as $N^3$, whereas it only scales quadratically in Eq.~\eqref{purityboundstateindependent} through the $\Gamma_{LM}$ [see Eq.~\eqref{offdiagrate}]. Therefore, our bound \eqref{purityboundstateindependent} is tighter than \eqref{QSL_Uzdin}, as can be seen in Fig.~\ref{fig:isotropic_R} showing the purity as a function of time for different $4$-qubit states.The \HOAP state is optimal, in the sense that it leads to the lowest purity at all times. This purity is, however, strictly larger than the bound \eqref{purityboundstateindependent}, which is therefore not tight.

We can also obtain a state-independent upper bound on the purity, valid for short and long times. Indeed, at short times, the purity is maximal when the opposite of the rate \eqref{ratet0} is the smallest, which occurs for coherent states (states for which $\mathcal{A}_1=0$). At long times, the multipoles that dominate the time dependence in Eq.~\eqref{puritytanisotropic} are for $L=1$ as they decay most slowly. From Eq.~\eqref{purityrhotTLM}, we know that $\sum_{M=0,\pm 1}|\rho_{1M}|^2$ is proportional to the purity of the one-qubit reduced density matrix $\rho_1$, which is also maximal for coherent states. Therefore, the purity on short and long time scales is upper bounded by the purity of coherent states $R_\mathrm{coh}(t)$ (for isotropic depolarization, all coherent states lead to the same purity). Using Eqs.~\eqref{rhoLMspinc} and \eqref{puritytanisotropic}, we get
\begin{equation}
	R_\mathrm{coh}(t) = \sum_{L=0}^N \frac{(2L+1)N!^2}{(N-L)!(N+L+1)!}\;e^{-2\gamma L(L+1)t}.
\end{equation}

\subsection{Entanglement dynamics} \label{sec:entanglementIsotropic}

\subsubsection{Entanglement survival time} 
Under isotropic depolarization, an $N$-qubit system initially in a pure entangled symmetric state $\ket{\psi}$ gradually loses its purity and entanglement over time. Asymptotically, the system reaches its unique stationary state, an equal weight mixture of all symmetric Dicke states, i.e., the maximally mixed state in the symmetric sector. This asymptotic state has the smallest purity and is separable because it can be written as a continuous convex mixture of symmetric separable states~\cite{2005Korbicz}. From Eq.~\eqref{TLMspincoh}, we indeed get
\begin{equation}
\rho_0 =\frac{1}{4 \pi} \int \left(|\Omega\rangle\langle \Omega|\right)^{\otimes N} d \Omega,
\end{equation}
where $|\Omega\rangle$ is a single-qubit state with Bloch vector pointing in the direction specified by the polar and azimuthal angles $\Omega\equiv (\theta, \varphi)$ and $d \Omega=\sin\theta d\theta d\varphi$. After a finite time of the evolution leading the system from its initial state $\ket{\psi}$ to $\rho_0$, the state becomes separable and remains so because the dynamics is separability preserving (see Appendix \ref{Appendix_coherencepreservation}). We call this particular time the entanglement survival time and denote it by $t_\mathrm{ES}$. After a longer but still finite time, the system's state enters a ball of absolutely separable states centered on $\rho_0$~\cite{2017Giraud}.
From this point on, any entanglement between the qubits that could potentially be retrieved by global unitary transformations on the system is permanently lost.

Our aim is to study the scaling of the entanglement survival time with the number of qubits, in particular for the most rapidly decohering states. However, in general, it is not practically feasible to compute this time, as this reduces to determining whether a mixed state is entangled or separable, a notoriously difficult problem. We will therefore use entanglement or separability criteria to give upper and lower bounds for $t_\mathrm{ES}$. As we shall see, these bounds follow the same scaling with $N$, which is consequently also the scaling of $t_\mathrm{ES}$ with $N$. For symmetric states, many distinct separability criteria (realignment criterion, spin-squeezing criterion, \ldots) are equivalent to the positive partial transpose (PPT) criterion~\cite{Tot09,Tot10}. In this work, we use the negativity to detect, and to some extent quantify, bipartite entanglement. Negativity is defined as the opposite of the sum of the negative eigenvalues of the partial transpose of $\rho$ with respect to some bipartition. A non-zero negativity indicates entanglement. We call the time when the negativity becomes zero the NPT (negative partial transpose) survival time $t_\mathrm{NPT}$. It is a lower bound for the entanglement survival time, i.e., $t_\mathrm{NPT}\leqslant t_\mathrm{ES}$. Symmetric $N$-qubit states are either genuinely entangled or fully separable~\cite{Eck02}. This implies that if the negativity is zero with respect to a certain bipartition $(q,N-q)$ of the qubits, but non-zero with respect to some other bipartition, the state is genuinely entangled and bound entangled with respect to the bipartition $(q,N-q)$. In this work, we will not investigate whether the states produced by isotropic depolarization are fully bound entangled~\cite{1998Horodecki}, but will focus on the distillable entanglement that is detected by the PPT criterion for the different bipartitions of the system, in the same spirit as the works~\cite{2004Dur,2005Dur}.

Even before using the PPT criterion, we can obtain a lower bound on the entanglement survival time for pure $q$-anticoherent initial states. We proceed as before by using the purity-based entanglement criterion which states that when $R(\rho)> R(\rho_q)$, the state $\rho$ is entangled. Then, a similar reasoning as the one leading to Eq.~\eqref{boundT} and taking into account the fact that $R(\rho_q)=1/(q+1)$ at any time because anticoherence is conserved now yields that $\rho(t)$ remains entangled for 
\begin{equation}\label{boundtES}
t < \frac{\ln \left(\frac{N(q+1)}{N-q}\right)}{2 \gamma N(N+1)}\leqslant t_\mathrm{ES}.
\end{equation}

\subsubsection{Initial negativity} 

For a pure state $\rho=\ket{\psi}\bra{\psi}$, the eigenvalues of the partial transpose $\rho^{T_q}$ of $\rho$ with respect to $q$ qubits are simple functions of the Schmidt coefficients $\sqrt{\lambda_i}$ of $|\psi\rangle$ for the bipartition $(q,N-q)$ (see, e.g.,~\cite{Nat18}). Equivalently, the eigenvalues of $\rho^{T_q}$ can be expressed in terms of the $q+1$ eigenvalues $\lambda_i$ of the reduced state $\rho_q=\tr_{\neg q}[\ket{\psi}\bra{\psi}]$. From there, a direct application of Lemma 1 of~\cite{Nat18} shows that the negativity of $|\psi\rangle$ is given by  (see also~\cite{Vid02})
\begin{equation}\label{negpurestate}
\mathcal{N}(|\psi\rangle)=\sum_{i>j}\sqrt{\lambda_i\lambda_j}.
\end{equation}
For GHZ states, the $q$-qubit reduced states read as $\rho_q= \mathrm{diag}(1/2,0,...,0,1/2)$ and the negativity is equal to $1/2$ for any bipartition. For Dicke states, the Schmidt decomposition~\cite{2003Stockton}
\begin{equation}
	|D_N^{(k)}\rangle = \sum_{\ell=0}^k \sqrt{\frac{\binom{q}{\ell}\binom{N-q}{k-\ell}}{\binom{N}{k}}}\, |D_q^{(\ell)}\rangle \otimes |D_{N-q}^{(k-\ell)}\rangle
\end{equation}
allows us to calculate the negativity across any bipartition $(q,N-q)$ through
\begin{equation}
	\mathcal{N}(|D_N^{(k)}\rangle ) = \tbinom{N}{k}\sum_{i=0}^q \sum_{j=0}^{i-1} \sqrt{\tbinom{q}{i} \tbinom{N-q}{k-i} \tbinom{q}{j} \tbinom{N-q}{k-j}}.
\end{equation}
In particular, we get for W states ($k=1$)
\begin{equation}
	\mathcal{N}(\ket{W}) = \frac{\sqrt{(N-q)q} }{N}.
\end{equation}

\begin{figure}
\begin{centering}
\includegraphics[width=0.45\textwidth]{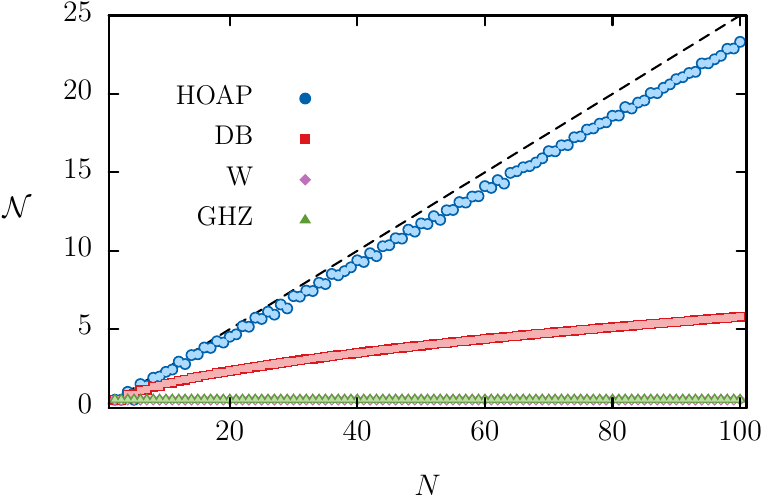}
\par\end{centering}
\caption{Negativity $\mathcal{N}$ with respect to the balanced bipartition $(\lfloor N/2\rfloor,\lceil N/2\rceil)$ of several families of initial pure states as a function of the number of qubits. We have $\mathcal{N}\approx 0.231\,N$ for \HOAP states (blue circles), $\mathcal{N}\approx -0.469+0.625\sqrt{N}$ for balanced Dicke states (DB, red squares), and $\mathcal{N}=1/2$ for W states (purple diamonds) and GHZ states (green triangles).
\label{fig:initial_entang}}
\end{figure}

In general, the maximal negativity $\mathcal{N}_\mathrm{max}=q/2$ is reached when all eigenvalues $\lambda_i$ are equal to $1/(q+1)$, i.e.\ when $\rho_q=\mathbb{1}/(q+1)$. The value $q/2$ is thus an upper bound for the negativity of pure states for a bipartition $(q,N-q)$. For an equal bipartition ($q=N/2$), the bound is equal to $N/4$ and thus scales linearly with the number of qubits. The bound is tight only when there are pure states $|\psi\rangle$ with maximally mixed $q$-qubit reductions, or, in other words, when $|\psi\rangle$ is $q$-anticoherent. In general, such states do not exist and the maximum possible order of anticoherence for pure $N$-qubit states is found to scale as $\sqrt{2N}$, much smaller than $N/2$ for large $N$~\cite{Bag15,ACstates}. There is also strong numerical evidence that pure anticoherent states of order $q=N/2$ only exist for $N=2,4$ and $6$~\cite{ACstates}. Figure~\ref{fig:initial_entang} shows the negativity with respect to a balanced bipartition for several types of initial states as a function of the number of qubits~\cite{footnotenegcomp}. We observe that for the \HOAP states found numerically in~\cite{ACstates}, we still find a linear scaling of the negativity with $N$, but with a prefactor of $0.231$ slightly smaller than the prefactor $1/4$ of the upper bound.

\subsubsection{NPT survival time scaling}

\begin{figure}[!h]
\begin{centering}
\includegraphics[width=0.47\textwidth]{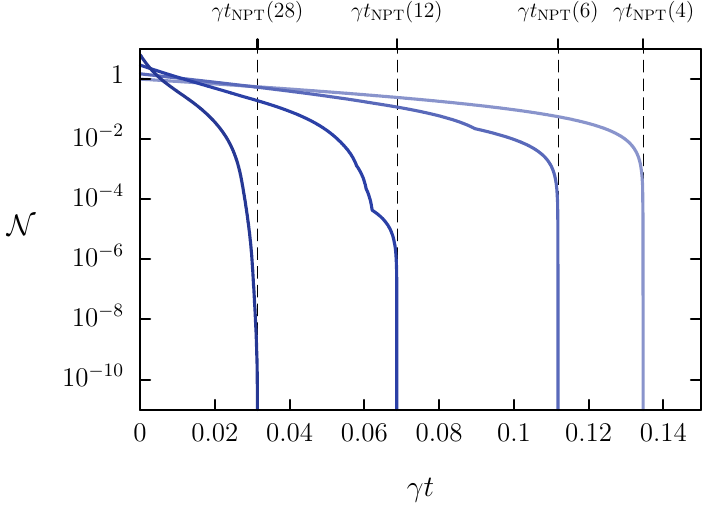}
\par\end{centering}
\caption{Negativity with respect to the $(\lfloor N/2\rfloor,\lceil N/2\rceil)$ bipartition of \HOAP states~\cite{ACstates}  for $N=4,6,12,28$ (from right to left) as a function of time for isotropic depolarization. The dashed lines indicate when the negativity drops to zero. \label{fig:entang_evolution_OAC}}
\end{figure}

In this subsection, we analyze the scaling of the NPT survival time with the number of qubits. To begin, Fig.~\ref{fig:entang_evolution_OAC} shows the negativity as a function of time for initial \HOAP states with different number of qubits. 
The logarithmic scale used in the plot clearly shows that the negativity drops to zero after a finite time, the NPT survival time $t_\mathrm{NPT}$, decreasing with the number of qubits~\cite{footnotetNPT}. We observe numerically that the balanced bipartition $(\lfloor N/2\rfloor,\lceil N/2\rceil)$ is, except for small numbers of qubits, the one for which $t_\mathrm{NPT}$ is the largest, while it is the smallest for the bipartition $(1,N-1)$. For this reason, in what follows, we focus only on these two bipartitions.

\begin{figure}[!h]
\begin{centering}
\includegraphics[width=0.48\textwidth]{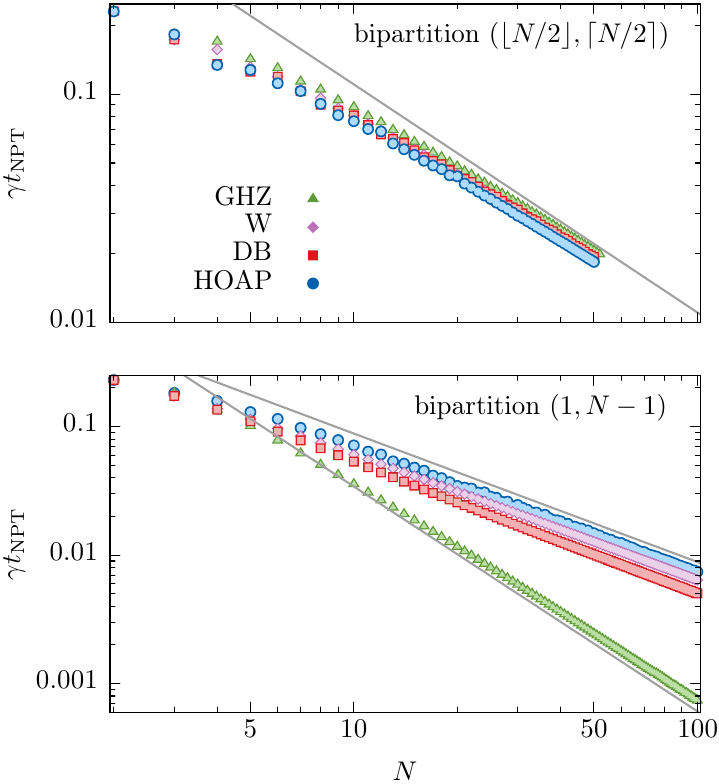}
\par\end{centering}
\caption{Negative partial transpose (NPT) survival time $t_{\mathrm{NPT}}$ as a function of $N$ for (top) the bipartition $(\lfloor N/2\rfloor,\lceil N/2\rceil)$, with the gray solid line showing the scaling $1/N$, and (bottom) the bipartition $(1,N-1)$, with the gray solid lines showing the scalings $1/N$ and $1/N^{1.75}$.
\label{fig:t_max_entang_N}}
\end{figure}

Figure \ref{fig:t_max_entang_N} (top panel) shows $t_\mathrm{NPT}$ for the balanced bipartition as a function of $N$. We see that $t_\mathrm{NPT}$ falls off as an inverse power law in $N$ for all states considered. The states that lose their entanglement the fastest and thus lead to the smallest value of $t_\mathrm{NPT}$ are the \HOAP states, even though they have the largest initial negativity. Similar data for the bipartition $(1,N-1)$ are displayed in the bottom panel of Fig.~\ref{fig:t_max_entang_N}. In this case, the GHZ states are the most fragile states under isotropic depolarization and we find that $t_\mathrm{NPT}$ falls off approximately as $N^{1.75}$. This observation can be contrasted with the fact that the entanglement in a GHZ state is known to be maximally fragile under the loss of a single qubit. Nevertheless, we see that the global entanglement is more robust than the local one and the optimal anticoherent states are found to be the first to become PPT with respect to any bipartition. Hence, anticoherent states are not only the most rapidly decohering states, but also the ones that become PPT the fastest.

\subsubsection{$P$-function separability time}

\begin{figure}[h!]
\begin{centering}
\includegraphics[width=0.475\textwidth]{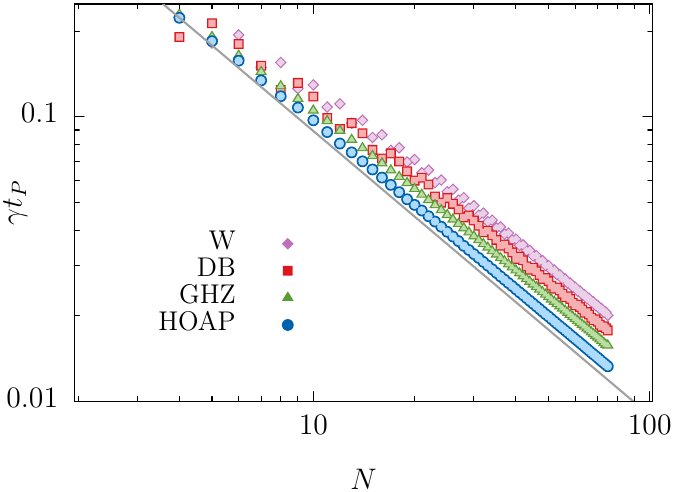}
\par\end{centering}
\caption{Shortest time after which the $P$ function \eqref{Pfunctime} becomes positive for any $\Omega$, ensuring the separability of the state. The gray solid line shows the scaling $1/N$.\label{fig:t_P}}
\end{figure}

For an initial pure entangled state $\rho=\ket{\psi}\bra{\psi}$, the $P$-function~\eqref{PfunctrhoLM} is necessarily not everywhere positive because otherwise the state would be separable. Over time, the state multipoles decay exponentially, at the exception of $\rho_{00}$. Therefore, after a finite time $t_P\geqslant t_{\mathrm{ES}}$, the $P$-function becomes everywhere positive (see Appendix \ref{Appendix_Pfunc}), which is a sufficient condition for separability. Because the dynamics is separability preserving, the state remains separable for all $t\geqslant t_P$. In Fig.~\ref{fig:t_P}, we show the time $t_P$ as a function of the number of qubits. For all the different types of states considered, we find an inverse power-law scaling of the form $t_P\sim 1/N$. It is for the \HOAP states that $t_P$ is the smallest and for W states  that $t_P$ is the largest for a given number of qubits.

\subsubsection{Time to reach a ball of absolutely separable state}

\begin{figure}[h!]
\begin{centering}
\includegraphics[width=0.475\textwidth]{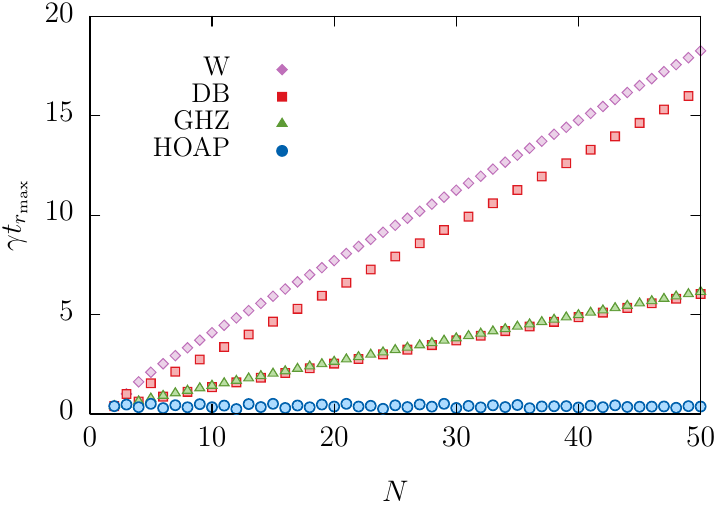}
\par\end{centering}
\caption{Time after which an initial pure spin-$j$ state enters the ball of radius \eqref{rmax} centered on the maximally mixed state as a result of depolarization, shown here for various states as a function of $N$. Note the difference in behavior for balanced Dicke states depending on the parity of $N$ (red squares).\label{fig:trmax}}
\end{figure}

We have computed the time it takes for an initial pure state to enter, as a result of depolarization, the ball of absolutely separable states of radius \eqref{rmax} centered on the maximally mixed state. Let us denote this time by $t_{r_{\mathrm{max}}}$. Figure \ref{fig:trmax} shows $t_{r_{\mathrm{max}}}$ as a function of $N$, the number of qubits. We can see that $t_{r_{\mathrm{max}}}$ increases linearly with $N$ for GHZ and Dicke states. In stark contrast, for highest-order anticoherent states, $t_{r_{\mathrm{max}}}$ is approximately independent of $N$, meaning that \HOAP states become absolutely separable after a time that depends only on the decoherence rates and not on the size of the system. More precisely, our high-precision numerics shows that $\gamma t_{r_{\mathrm{max}}}\approx 0.384$ for \HOAP states, $\gamma t_{r_{\mathrm{max}}}\approx 0.116 (N+2)$ for GHZ and balanced Dicke states with even $N$, $\gamma t_{r_{\mathrm{max}}}\approx 0.341 (N-2)$ for balanced Dicke states with odd $N$, and $\gamma t_{r_{\mathrm{max}}}\approx 0.35 N+0.81$ for W states.

Our results are summarized in Table~\ref{tabentang}. It is interesting to compare them with those obtained for an ensemble of qubits subjected to \emph{individual} decoherence. In that case, it has been shown that the time $t_{\mathrm{NPT}}$ calculated for the bipartition $(\lfloor N/2\rfloor,\lceil N/2\rceil)$ generally increases with $N$, notably for GHZ states~\cite{2004Dur, 2005Dur, 2002Kempe, Aol08, 2015Aolita}. It was therefore concluded that the entanglement survival time $t_{\mathrm{ES}}$ also increases with $N$. Clearly, these conclusions are no longer valid for \emph{collective} depolarization since we have just shown that then both times decrease with $N$. Highly entangled symmetric states thus appear to be more fragile to collective depolarization than to individual depolarization, which we attribute to superdecoherence that is absent for individual depolarization.

\begin{table}[h!]
\begin{centering}
\begin{tabular}{c||c|c|c|c|c}
 & $\mathcal{N}$ & $t_{\mathrm{NPT}}$ & $t_{\mathrm{ES}}$ & $t_{P}$ & $t_{r_{\mathrm{max}}}$ \\
\hline \hline 
GHZ & $1/2$ & $\sim 1/N$ & \textcolor{orange}{$\sim 1/N$} & $\sim 1/N$ & $\sim N$ \\
\hline
W & $1/2$ & $\sim 1/N$ & \textcolor{orange}{$\sim 1/N$} & $\sim 1/N$ & $\sim N$ \\
\hline
DB & $\sim \sqrt{N}$ & $\sim 1/N$ & \textcolor{orange}{$\sim 1/N$} & $\sim 1/N$ & $\sim N$ \\
\hline
\HOAP & $\sim N$ & $\sim 1/N$ & \textcolor{orange}{$\sim 1/N$} & $\sim 1/N$ & const \\
\end{tabular}
\caption{Scaling laws with the number of qubits of the initial negativity and the different characteristic times defined in the text for several families of states subject to isotropic depolarization. The scaling laws for $t_\mathrm{ES}$ (highlighted in orange) have been deduced from the inequalities $t_\mathrm{NPT}\leqslant t_\mathrm{ES} \leqslant t_P$ and the scaling laws for $t_\mathrm{NPT}$ and $t_P$.
\label{tabentang}}
\par\end{centering}
\end{table}

\subsubsection{Dynamical generation of bound entangled states}

\begin{figure*}[t!]
\begin{centering}
\begin{overpic}[scale=0.3]{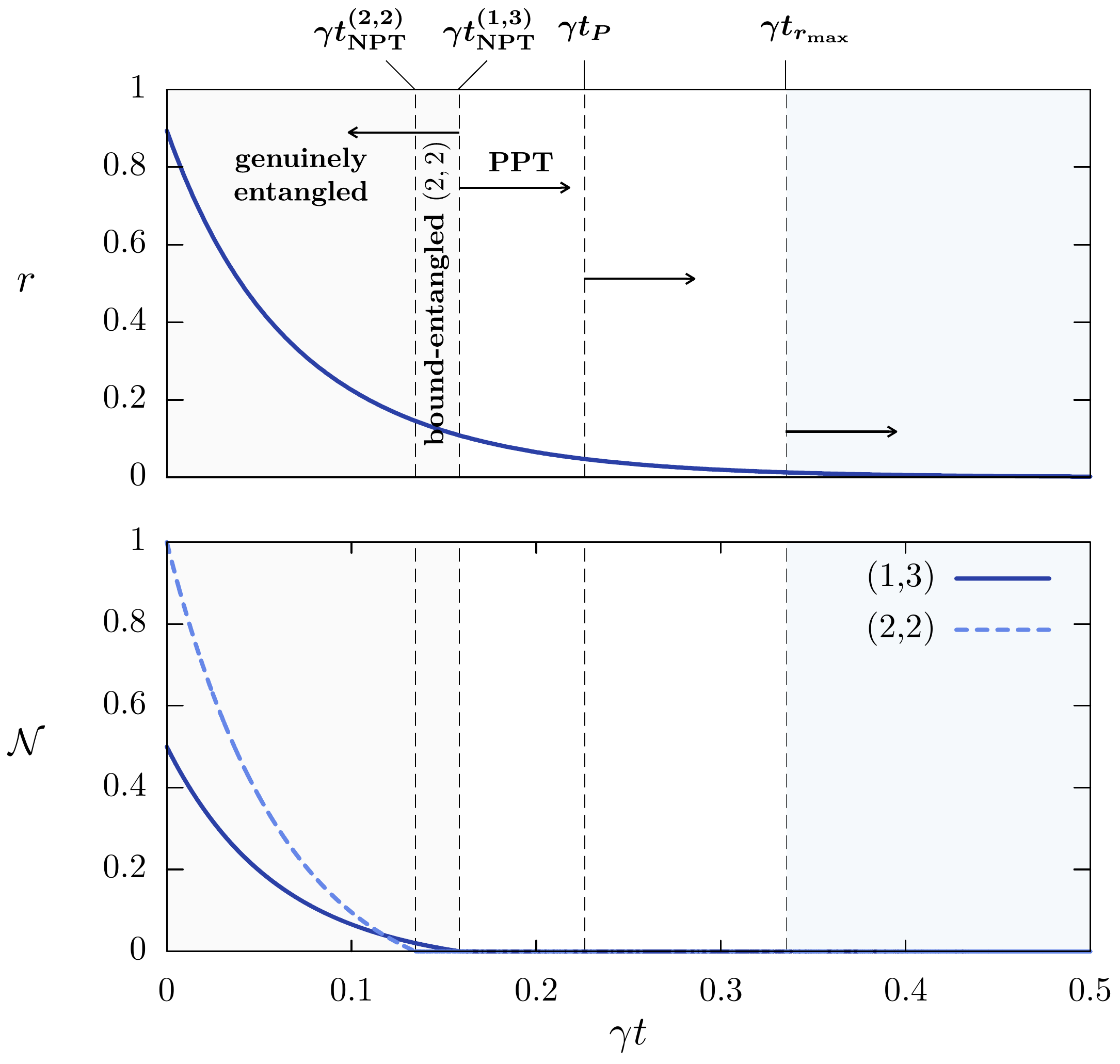}
\put(47,77.5){\rotatebox{0}{\scriptsize \parbox[t]{3cm}{
\centering \textbf{positive\\ $P$-function\\ \eqref{PfunctrhoLM}}}}}
\put(69,67){\rotatebox{0}{\scriptsize \parbox[t]{3cm}{
\centering \textbf{$\in$ ball of\\ absolutely\\ separable states\\ of radius \eqref{rmax}}
}}}
\end{overpic}
\caption{Top: Distance to the maximally mixed state as a function of time when starting from the $4$-qubit pure state \eqref{eq:anticoherent_state}. Bottom: Negativity as a function of time for the two possible bipartitions $(1,3)$ and $(2,2)$. The times $t_\mathrm{NPT}$, $t_P$ and $t_{r_{\mathrm{max}}}$ introduced in the text are indicated on top of the figure (with the bipartition as superscript for $t_\mathrm{NPT}$). It holds that $t_\mathrm{NPT}\leqslant t_\mathrm{ES} \leqslant t_P\leqslant t_{r_{\mathrm{max}}}$. \label{fig:isotropic_N4}}
\end{centering}
\end{figure*}

We now focus in more detail on the case of a few qubits for which we can obtain further analytical results. To begin with, let us consider a $4$-qubit system initially in the pure anticoherent state to order 2:
\begin{equation}
|\psi_\mathrm{tet}\rangle=\frac{1}{2}\left(|D_4^{(0)}\rangle +i\sqrt{2}\,|D_4^{(2)}\rangle +|D_4^{(4)}\rangle \right).\label{eq:anticoherent_state}
\end{equation}
The time-evolved state $\rho(t)$, which follows from Eqs.~\eqref{gensoldiag} and \eqref{gensoldiagjm}, is represented in the Dicke basis $\{|D_4^{(k)}\rangle: k=0,\ldots,4\}$ by the matrix
\begin{equation}\label{stateN4}
\rho(t)=\left(
\begin{array}{ccccc}
 \frac{u+1}{5} & 0 & -\frac{i u^{3/5}}{2^{3/10}} & 0 & u \\
 0 & \frac{1-4 u}{5} & 0 & 0 & 0 \\
 \frac{i u^{3/5}}{2^{3/10}} & 0 & \frac{6 u+1}{5} & 0 & \frac{i u^{3/5}}{2^{3/10}} \\
 0 & 0 & 0 & \frac{1-4 u}{5} & 0 \\
 u & 0 & -\frac{i u^{3/5}}{2^{3/10}} & 0 & \frac{u+1}{5} \\
\end{array}
\right),
\end{equation}
where $u=u(t)=e^{-20 t}/4$. The partial transpose of \eqref{stateN4} has only one eigenvalue that can potentially be negative, respectively denoted by $\lambda_-^{(1,3)}$ for the bipartition $(1,3)$ and $\lambda_-^{(2,2)}$ for the bipartition $(2,2)$. These eigenvalues are given by
\begin{equation}
\lambda_-^{(1,3)}(t)=\frac{e^{-20 t} }{20} \left(e^{20 t}-5\, e^{8 t}-6\right)
\end{equation}
and
\begin{equation}
\begin{aligned}
\lambda_-^{(2,2)}(t) & = -\frac{e^{-20 t} }{30} \Big( 3-3\, e^{20 t}\\
& \quad +\sqrt{e^{16 t} \left[4\, e^{4 t} \left(e^{20 t}+3\right)+75\right]+9}\,\Big).
\end{aligned}
\end{equation}
The eigenvalue $\lambda_-^{(2,2)}(t)$ cancels at a time $\gamma t_{\mathrm{NPT}}^{(2,2)}\approx 0.134582$, less than $\gamma t_{\mathrm{NPT}}^{(1,3)}\approx 0.158384$, the time at which $\lambda_-^{(3,1)}(t)$ cancels. Therefore, there is a time interval during which the negativity is zero for the cut $(2,2)$ while it is non-zero for the cut $(1,3)$. In this time interval, the state generated by isotropic depolarization is bound entangled with respect to the $(2,2)$ bipartition, and thereby realizes a $3\times 3$ bipartite symmetric bound entangled state~\cite{Tot09}. Figure~\ref{fig:isotropic_N4} illustrates this dynamical evolution of entanglement under collective depolarization in this particular case. It should be noted that for the initial GHZ, W and Dicke states, no such bound entanglement is produced during the dynamics. Similarly, we found that other types of bound entangled states can be generated by isotropic depolarization starting from pure anticoherent states. For example, starting from the $5$-qubit and $6$-qubit anticoherent state given in Table~\ref{tabACstates}, isotropic depolarization leads to bound entangled states with respect to the $(3,2)$ bipartition~\cite{BE5o2} and $(3,3)$ bipartition~\cite{BE3}.

\section{Anisotropic depolarization}
\label{Sec:anisotrop}

In this section, we turn to anisotropic depolarization, where the rates in the three spatial directions are not equal, i.e., $\gamma_{x}=\gamma_{y}=\gamma_{\perp z}\ne\gamma_{z}$. We start in Sec.~\ref{sec:pureDephasing} with the well-known case of pure dephasing ($\gamma_{\perp z}=0$, $\gamma_{z}\ne 0$), which is by far the most explored type of decoherence, mainly because a corresponding microscopic model can be solved exactly~\cite{1996Palma,2002Reina}. We show how our approach allows us to easily recover results from the literature. Then we turn in Sec.~\ref{sec:purityLossAnisotropic} to the more interesting case of anisotropic depolarization ($\gamma_{\perp z}\ne\gamma_z$) and look in Sec.~\ref{purityLossOptimization} for the initial states with the largest purity decay rate. The minimization of the purity at longer times is addressed for small quantum spin numbers in Sec.~\ref{sec:purityMinimization}. Finally, the dynamics of entanglement is briefly discussed in Sec.~\ref{sec:entanglementAnisotropic}.

\subsection{Pure dephasing} \label{sec:pureDephasing}

One of the most studied types of decoherence is pure dephasing ($\gamma_{\perp z}=0,\gamma_z\ne 0$)~\cite{Li07, 2008Guhne, 2019Korbicz}. The general solution \eqref{gensoldiag} reads as in this case $\rho_{LM}(t)=e^{-(M^{2}\gamma_z+i\omega M) t}\rho_{LM}(0)$. From Eq.~\eqref{gensoldiagjm} for the matrix elements $\rho_{mm'}$ in the Dicke basis and the fact that the Clebsch-Gordan coefficients are non vanishing only when $M=m'-m$, we recover the results of~\cite{1996Palma,2002Reina}, namely, that the rate of dephasing is given by $M^{2}\gamma_z= (m'-m)^{2}\gamma_z$. The higher the Hamming distance $m'-m$, the faster the decay of the off-diagonal density matrix element in the Dicke basis. Only the populations remain constant. Moreover, the dephasing rate increases quadratically with the Hamming distance. This is in contrast to the case of independent, rather than collective, dephasing, where the dephasing rate increases only linearly with $m'-m$. The pure states that reach their stationary state the fastest, i.e., those with the highest decoherence rate, are those for which the only non-vanishing coherences have the largest possible Hamming distance. These states are the $N$-qubit GHZ states $|\mathrm{GHZ}\rangle=\frac{1}{\sqrt{2}}(|D_N^{(0)}\rangle +|D_N^{(N)}\rangle)$. Their non-zero state multipoles are $\rho_{N\pm N}=(\mp 1)^{N}/2$ and, for even $L$ only,
\begin{equation}
\rho_{L0}=\sqrt{\tfrac{(2L+1)(N!)^2}{(N-L)!(N+L+1)!}},
\end{equation}
so that the purity decays at a constant rate
\begin{equation}
\frac{\dpuritydt}{\purity}  = -2\gamma_{z} \,\overline{M^2} = -2\gamma_{z}N^2.
\end{equation}
Note that the W and balanced Dicke states belong to the decoherence-free subspace of pure dephasing~\cite{2003Lidar}.

\subsection{Purity loss rate} \label{sec:purityLossAnisotropic}

In the more general case of anisotropic depolarization ($\gamma_{\perp z}\ne\gamma_z$), the rate of change of the purity is given by
\begin{equation}\label{Rdotanisotropic}
	\dpuritydt=-2\sum_{L,M} \Big(M^2\gamma_z + [L(L+1)-M^{2}]\gamma_{\perp z}\Big)|\rho_{LM}|^2.
\end{equation}
For a pure state $\ket{\psi}$, $\dpuritydt$ can be expressed only in terms of variances of the spin components in state $\ket{\psi}$. Indeed, for pure states, the following relations hold (see Appendix \ref{Appendix_PureDecoRate}):
\begin{equation}\label{varJpsi}
\begin{aligned}
& \Delta J_z^2 = \frac{1}{2}\sum_{L,M}  M^2\, |\rho_{LM}|^2, \\
& \Delta J_x^2+\Delta J_y^2 = \frac{1}{2}\sum_{L,M} \left[L(L+1)-M^2\right]\, |\rho_{LM}|^2
\end{aligned}
\end{equation}
that can be combined with Eq.~\eqref{Rdotanisotropic} to obtain
\begin{equation}
	\dpuritydt_{\ket{\psi}}=-4 \left[\gamma_z\,\Delta J_z^2+ \gamma_{\perp z}\,\big(\Delta J_x^2+\Delta J_y^2\big)\right].
	\label{eq:purity_variances}
\end{equation}
The general result~\eqref{eq:purity_variances}, which is valid for any spin, suggests controlling depolarization on short time scales by appropriately squeezing the initial state, i.e., making $\Delta J_z^2$ smaller or larger than $\Delta J_x^2+\Delta J_y^2$, depending on the value of the depolarization rates $\gamma_z$ and $\gamma_{\perp z}$. Interestingly, a similar result applies to continuous variables quantum states~\cite{Ser04}. In particular, for macroscopic quantum states, a slowing down of decoherence due to photon loss has been experimentally demonstrated via squeezing in phase space~\cite{2018Jeannic, 2018Brewster}.

\subsection{Optimization of $\dpuritydt_{\ket{\psi}}$}\label{purityLossOptimization}

In the following, we show how to find the pure states that maximize (the opposite of) the purity loss rate~\eqref{eq:purity_variances}. First, let us remark that the variances of the spin components verify
\begin{equation} \label{sum_variances}
\begin{aligned}
\Delta J_x^2+\Delta J_y^2+\Delta J_z^2 ={}& \tfrac{N}{2}\left(\tfrac{N}{2}+1\right)-|\langle \mathbf{J}\rangle|^2
\end{aligned}
\end{equation}
because $\mathbf{J}^2=\frac{N}{2}\left(\frac{N}{2}+1\right)\mathbb{1}$, and
\begin{equation}\label{varJalpha}
0\leqslant \Delta J_\alpha^2= \langle J_\alpha^2\rangle-\langle J_\alpha\rangle^2\leqslant \langle J_\alpha^2\rangle \leqslant \frac{N^2}{4},
\end{equation}
where the last inequality comes from the fact that the largest eigenvalue of $J_\alpha^2$ is equal to $N/2$.

\subsubsection{Case I: $\gamma_z>\gamma_{\perp z}$}
In this case, the decoherence rate~\eqref{eq:purity_variances} is maximized by 
maximizing $\Delta J_z^2$. Let us show that the upper bound for the variance given in Eq.~\eqref{varJalpha}, equal to $N^2/4$, can be reached. It can only be reached when $\langle J_z \rangle = 0$. For a pure state $\ket{\psi}$ with expansion in the Dicke basis
\begin{equation}
\ket{\psi}=\sum_{k=0}^Nd_k\,\ket{D_N^{(k)}},\quad d_k\in\mathbb{C},
\end{equation}
the condition $\langle J_z\rangle =0$ reads as
\begin{equation}\label{AC1dk}
\sum_{k=0}^N|d_{k}|^2(N-2k)=0,
\end{equation}
and the variance $\Delta J_z^2$ then reduces to
\begin{equation}
	 \langle J_z^2\rangle = \frac{N^2}{4}+\sum_{k=0}^N |d_{N-k}|^2\,k(k-N).
	\label{JzsquaredDicke}
\end{equation}
The latter expression is equal to $N^2/4$ when the sum on the right-hand side vanishes, which occurs when all Dicke coefficients $d_k$ are equal to zero, except $d_0$ and $d_N$. But, then, Eq.~\eqref{AC1dk} gives $|d_0|^2=|d_N|^2$ directly. The corresponding normalized state is the GHZ state, for which Eq.~\eqref{eq:purity_variances} yields the maximal purity loss rate for $\gamma_z>\gamma_{\perp z}$
\begin{equation}
	\dpuritydt_{\ket{\mathrm{GHZ}}}=- 2\gamma_{\perp z}N - \gamma_z N^2.
\end{equation}

\subsubsection{Case II: $\gamma_z<\gamma_{\perp z}$}

In this case, the decoherence rate~\eqref{eq:purity_variances} is maximized by 
maximizing $\Delta J_x^2+\Delta J_y^2$. Rewriting Eq.~\eqref{sum_variances} as
\begin{equation}
	\Delta J_x^2+\Delta J_y^2 =  \tfrac{N}{2}\left(\tfrac{N}{2}+1\right) - \langle J_x\rangle^2 - \langle J_y \rangle^2 - \langle J_z^2\rangle,
\end{equation}
we see that a state verifying $\langle J_x\rangle = \langle J_y \rangle = 0$ and,  at the same time, $\langle J_z^2\rangle=0$ is optimal. For even $N$, $\langle J_z^2\rangle$ given by Eq.~\eqref{JzsquaredDicke} is equal to zero when $|d_k|^2=1$ for $k=N/2$ and $|d_k|^2=0$ otherwise. The corresponding state is the balanced Dicke state, which also verifies $\langle J_x\rangle = \langle J_y \rangle = 0$ and is thus optimal. Similarly, for odd $N$, we find that the optimal state has $|d_{k}|^2=1$ either for $k=(N+1)/2$ or $k=(N-1)/2$ and $|d_{k}|^2=0$ otherwise. In the end, the maximal purity loss rate for $\gamma_z<\gamma_{\perp z}$ is given by
\begin{equation}
	\dpuritydt_{\ket{\mathrm{DB}}} = \left\{\begin{array}{ll}
	-\gamma_{\perp z} N (N+2), & \mathrm{even}~N\\
	-\gamma_{\perp z} [N(N+2)-1], & \mathrm{odd}~N
	\end{array}\right.
\end{equation}
and is independent of $\gamma_z$.

\subsection{Minimization of purity at any fixed time}\label{sec:purityMinimization}

\begin{figure}[h!]
\begin{centering}
\includegraphics[width=0.45\textwidth]{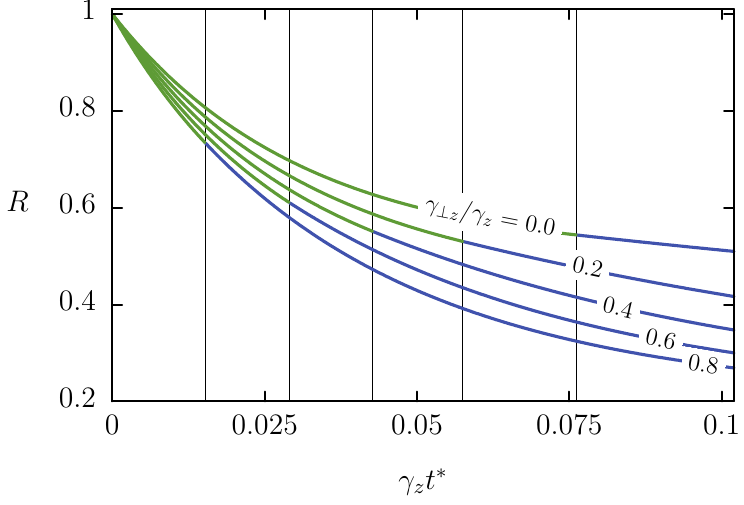}
\par\end{centering}
\caption{Minimum achievable purity at time $t^*$ for different values of $\gamma_{\perp z}/\gamma_z <1$. The optimal initial states are the GHZ state (green curves) at short times and the $\ket{\mu^*}$ states given by Eq.~\eqref{eq:optimal_state_N=4} (gray curves) at longer times. Vertical dashed lines locate the transition from GHZ to $\ket{\mu^*}$ which occurs when $\mathrm{Im}(\mu^*)$ vanishes. \label{fig:N=4_z>xy}}
\end{figure}

Until now, we have identified the states which decohere the fastest \emph{at short times} for any possible value of the rates $\gamma_{\perp z}$ and $\gamma_z$. It is much more difficult to answer the question of which initial pure states give the smallest purity after an arbitrary period of time $t^*$. Through a combination of analytical and numerical methods, we were able to answer this question for 2 and 4 qubits. 
First, we have observed that, for $N=2$ and $4$ (but not $N=3$), the optimal initial states found numerically are always $1$-anticoherent. Taking this observation as a working hypothesis, we can perform the optimization analytically. To this end, let us denote by $R_{\ket{\psi}}(t^*)$ the purity of the system's state at time $t^*$ when the system is initially in state $\ket{\psi}$. We then define the derivative of this purity with respect to a pure state $|\phi\rangle$ as
\begin{equation}\label{derivative}
\frac{dR_{\ket{\psi}}(t^*)}{d|\phi\rangle} = \lim\limits_{\epsilon \to 0} \frac{R_{\ket{\psi_\epsilon}}(t^*)-R_{\ket{\psi}}(t^*)}{\epsilon},
\end{equation}
where $|\psi_\epsilon\rangle $ is the normalized state 
\begin{equation}
|\psi_\epsilon\rangle = \mathcal{N}(|\psi\rangle+\epsilon|\phi\rangle).
\end{equation}
In order to find optimal states, we look for critical states that cancel the derivative \eqref{derivative} for a fixed value of $t^*$. 

\subsubsection{$j=1$ or $N = 2$ qubits}
For $N=2$, the only critical states are the GHZ and balanced Dicke states, whatever the value of $t^*$. They are both $1$-anticoherent~\cite{Bag15}. The GHZ and balanced Dicke states were previously found to be optimal at short times for $\gamma_z > \gamma_{\perp z}$ and $\gamma_z<
\gamma_{\perp z}$, respectively. Our numerics shows that this is also the case for longer times, and thus for any $t^*$.

\begin{figure}[h!]
\begin{centering}
\includegraphics[width=0.45\textwidth]{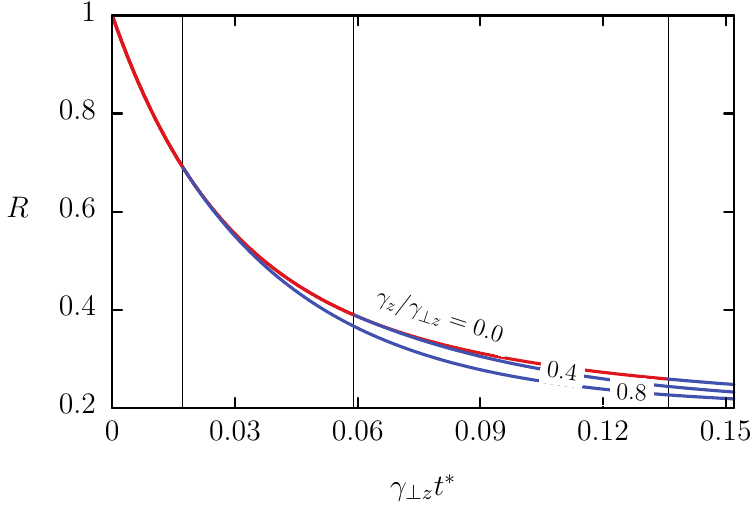}
\par\end{centering}
\caption{Same as for Fig.~\ref{fig:N=4_z>xy} but for different values of $\gamma_{z}/\gamma_{\perp z}<1$. The optimal initial states are the balanced Dicke state (red curve) at short times and the $\ket{\mu^*}$ states given by Eq.~\eqref{eq:optimal_state_N=4} (grey curves) at longer times. \label{fig:N=4_xy>z}}
\end{figure}

\subsubsection{$j=2$ or $N = 4$ qubits}

\begin{figure}
\begin{centering}
\includegraphics[width=0.45\textwidth]{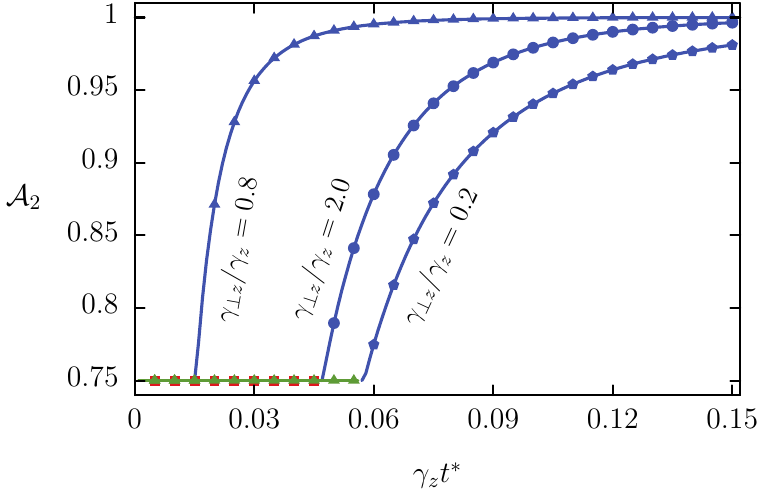}
\par\end{centering}
\caption{Measure of anticoherence to order $2$ of optimal states (see text) as a function of the final time $t^*$ for $N=4$ and different values of the depolarization rates. Symbols correspond to data obtained by numerical optimization and solid curves show the analytical predictions based on Eqs.~\eqref{mustate} and \eqref{eq:optimal_state_N=4}. \label{fig:ACMeas_N=4}}
\end{figure}

We assume that the optimal initial $4$-qubit states are all $1$-anticoherent, which our numerics confirms, regardless of the final time $t^*$ and the depolarization rates. In~\cite{Bag15}, it was shown that any $1$-anticoherent $4$-qubit state can be brought by rotation to the form
\begin{equation}\label{mustate}
	|\mu\rangle =\mathcal{N}\big(|D_4^{(0)}\rangle+\mu |D_4^{(2)}\rangle+|D_4^{(4)}\rangle\big)
\end{equation}
with $\mu\in\mathbb{C}$ and $\mathcal{N}$ a normalization constant. We can always find a critical state of the form~\eqref{mustate}. Indeed, by canceling the derivative of the purity at time $t^*$ for the initial state \eqref{mustate} and solve for the real and imaginary parts of $\mu$, we find a critical value $\mu^*$ with $\mathrm{Re}(\mu^*)=0$ and
\begin{equation}
\mathrm{Im}(\mu^*) = \sqrt{\tfrac{7 \left(e^{8 (4 \gamma_{\perp z}-3 \gamma_z)t^*}-e^{8\gamma_z t^*}\right)}{4\,e^{4 (7 \gamma_{\perp z}+2 \gamma_z)t^*}-7\,e^{24\gamma_{\perp z} t^*}+3\, e^{8 \gamma_z t^*}}+2}.
\label{eq:optimal_state_N=4}
\end{equation}
Let us make some observations on this result. First, when $\gamma_{\perp z}=\gamma_z$, Eq.~\eqref{eq:optimal_state_N=4} reduces to $\mu^*=i\sqrt{2}$ and Eq.~\eqref{ACR} then gives $\mathcal{A}_2=1$, so that we recover the result of the previous section that, for isotropic depolarization, the optimal state is at any time a $2$-anticoherent state. Next, the critical value \eqref{eq:optimal_state_N=4} exists only for times for which the argument in the square root is non-negative, i.e., for $t^*>t^*_{\mu^*}$ with $t^*_{\mu^*}$ a time that depends only on the rates. 

We show in Fig.~\ref{fig:N=4_z>xy} the evolution of the purity for the optimal state $\ket{\mu^*}$ for different values of $\gamma_{\perp z}/\gamma_{z}$ with $\gamma_{z}>\gamma_{\perp z}$. The vertical dashed lines, drawn at $t^*=t^*_{\mu^*}$, indicate the time at which the initial optimal state changes from the GHZ state to $|\mu^*\rangle$. Finally, we note that $\mu^*\to i\sqrt{2}$ when $t\to\infty$, meaning that the asymptotically optimal state always converges to the pure anticoherent state to order 2 given in Eq.~\eqref{eq:anticoherent_state}.
The same type of behavior can be seen in Fig.~\ref{fig:N=4_xy>z}, where we let $\gamma_z/\gamma_{\perp z}$ vary with $\gamma_{z}<\gamma_{\perp z}$. Here, the optimal state changes from the balanced Dicke state to $\ket{\mu^*}$. As all non-zero state multipoles of the balanced Dicke state have $M=0$, the purity decay is insensitive to the value of the rate $\gamma_z$.

Finally, in Fig.~\ref{fig:ACMeas_N=4}, we compare the rotationally invariant measure of $2$-anticoherence of the numerically obtained optimal states and the $\ket{\mu^*}$ states as a function of $t^*$. As can be seen, both give the same results. At short times, $\mathcal{A}_2$ is constant and equal to $3/4$. The corresponding state is either the GHZ state when $\gamma_z<\gamma_{\perp z}$, or the balanced Dicke state when $\gamma_z>\gamma_{\perp z}$. At longer times, $\mathcal{A}_2$ increases monotonically with $t^*$ and tends to $1$ as $t^*\to\infty$. All this shows that the optimization results of the purity for arbitrary times are already complex for a small number of qubits.

\begin{widetext}

\begin{figure*}[b]
\begin{centering}
\includegraphics[width=1.0\textwidth]{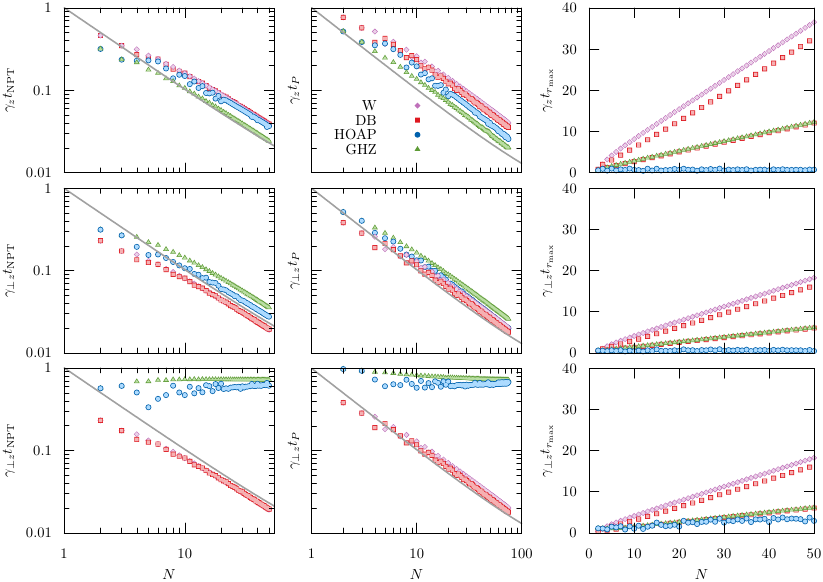}
\par\end{centering}
\caption{Evolution of the different characteristic times [NPT survival time $t_\mathrm{NPT}$, $P$-function separability time $t_P$, and time $t_{r_{\mathrm{max}}}$ to reach the ball of absolutely separable state of radius \eqref{rmax}] as a function of the number of qubits for W states (purple), balanced Dicke states (red), GHZ states (green), and \HOAP states (blue). Top: $\frac{\gamma_{\perp z}}{\gamma_z}=0.5$. Middle: $\frac{\gamma_z}{\gamma_{\perp z}}=0.5$. Bottom: $\frac{\gamma_z}{\gamma_{\perp z}}=0.0$.\label{fig:anisotropic_panels}}
\end{figure*}

\end{widetext}

\subsection{Entanglement dynamics}\label{sec:entanglementAnisotropic}

As for isotropic depolarization, we now look at the evolution with the number of qubits of the different characteristic times related to entanglement. Our numerical results are displayed in Fig.~\ref{fig:anisotropic_panels}. The first two rows show that, when $\gamma_z\ne 0$, the times $t_\mathrm{NPT}$ and $t_P$ (and thus also the entanglement survival time) follow the same scaling $\sim 1/N$ as for isotropic depolarization for all state families considered in this work. However, when $\gamma_z=0$, we observe that these times increase or decrease only very slightly with $N$ for GHZ and \HOAP states (see left two bottom panels) for $N\lesssim 50$. As for the time $t_{r_{\mathrm{max}}}$, we can see in the right column that it is the smallest for the \HOAP states for all the values of the rates considered. This suggests that \HOAP states are, also for anisotropic depolarization, the ones that lead to the smallest purity at long times, in line with our detailed analysis of the $4$-qubit case in the preceding subsection. An in-depth study of these behaviours as a function of the depolarization rates would be desirable, but is beyond the scope of this work.

\section{Conclusion}

In this work, we presented a general study of the depolarization dynamics of an arbitrary spin using tools from quantum information theory. Considering that the single spin-$j$ results from the coupling of $N=2j$ constituent spin-$1/2$ or qubits, we analyzed depolarization in terms of multiqubit states evolving in the symmetric sector of the full Hilbert space. A simple analytical solution to the general master equation \eqref{MEq} was obtained in the multipole operator basis and exploited to find the states featuring the most rapid decoherence, both for isotropic and anisotropic depolarization.

In the case of isotropic depolarization, we proved that pure- or mixed-state entanglement between the constituent qubits is a necessary condition for superdecoherence to occur, in stark contrast to the closely related phenomenon of superradiance which can occur without any entanglement~\cite{Wol14}. Our result suggests that superdecoherence could be used as an entanglement witness for multiqubit symmetric states, or equivalently as a non-classicality witness for spin states. We found a relationship [Eq.~\eqref{ratet0}] between the initial decoherence rate of pure states and their measure of anticoherence (and thus their linear entanglement entropy). This allowed us to identify a class of maximally entangled pure states distinct from the GHZ and Dicke states, known as anticoherent states, which exhibit the highest initial decoherence rates and lead to the states with the lowest purity at any time, starting from a pure state. The potential of these anticoherent spin states for rotation sensing has been demonstrated in a series of works (see, e.g., \cite{Gol18, Mar20, Gol21}). The entanglement of the different families of states listed in Table~\ref{tabstates} and its evolution over time was studied numerically. More precisely, the entanglement survival time, that we lower and upper bounded using the PPT entanglement criterion and a sufficient separability criterion based on the $P$ function, was shown to scale as $1/N$ with $N$ the number of qubits, consistently with the bound \eqref{boundtES}. Although other families of states show the same scaling (see Table~\ref{tabentang}), we found the entanglement of anticoherent states to be more fragile to depolarization than for GHZ, W or even Dicke states. The $t_{\mathrm{ES}}\sim 1/N$ scaling we found is also different from that of individual, rather than collective, decoherence. We attribute this difference to superdecoherence that is absent for individual depolarization. A detailed analysis for a few qubits showed that isotropic depolarization can lead to the dynamical creation of bound entanglement across balanced bipartitions (see, e.g., Fig.~\ref{fig:isotropic_N4}). Then, we studied when a state enters the ball of absolutely separable states of radius \eqref{rmax}. The states belonging to this ball are too mixed for any unitary operation to create entanglement. It was found that pure states with maximum order of anticoherence enter this ball after a time roughly independent of the number of qubits, in sharp contrast to other states for which this time increases linearly with $N$. This again points to the extreme fragility of entanglement in anticoherent states against depolarization.

In the case of anisotropic depolarization, we first related the initial purity loss rate to the variances of the spin components in a pure state [Eq.~\eqref{eq:purity_variances}]. We then showed for any spin that the maximum purity loss rate at short times is achieved by the GHZ states for $\gamma_z>\gamma_{\perp z}$ and by the balanced Dicke state for $\gamma_z<\gamma_{\perp z}$. As for the dynamics at longer times, we have completely identified the pure states of $N=2$ and $4$ qubits that display the lowest purity after an arbitrary fixed time. The extremal states for $N=4$ were found to exhibit a transition from the GHZ/balanced Dicke state to a parametric anticoherent state given by Eq.~\eqref{mustate}. 

Regarding the perspectives of this work, the entanglement dynamics for anisotropic depolarization was only touched upon and deserves further investigation. In particular, the question remains whether entanglement between the constituent qubits is a necessary condition for superdecoherence to occur in this case. Another perspective this work suggests is to analyze the potential of anticoherent spin states, which we have shown to be the most sensitive to isotropic depolarization, for quantum parameter estimation and quantum sensing strategies based on decoherence, in line with the works~\cite{2014Rozema,2020Gebia, Kuk21}. Finally, recent experiments with multiphoton quantum states~\cite{2014Rozema} or the electron spin of dysprosium atoms whose ground state has an angular momentum $j=8$~\cite{2021Satoor} suggests that some of the predictions made in this work can already be verified in the laboratory.

\begin{acknowledgments}
Computational resources were provided by the Consortium des Equipements de Calcul Intensif (CECI), funded by the Fonds de la Recherche Scientifique de Belgique (F.R.S.-FNRS) under Grant No. 2.5020.11. Most of the computations were done with the Julia programming language, in particular using the DifferentialEquations.jl package~\cite{Rac17}. We thank François Damanet for comments on the manuscript.
\end{acknowledgments}

\clearpage

\begin{widetext}
\section*{Examples of pure anticoherent states in the \basis}

We give in Table~\ref{tabACstates} below examples of pure states with the maximum achievable order of anticoherence (taken from Refs.~\cite{2006Zimba,Bag15,Bjo15,ACstates}) expressed in both the standard angular momentum basis and the multipole operator basis. These states are abbreviated as \HOAP states in this work. 

\setlength{\tabcolsep}{5pt}
\renewcommand{\arraystretch}{1.5}

\begin{table}[h]
\begin{centering}
\begin{tabular}{|c|c|@{\hskip 10pt}l|}
\hline 
$j$ & $q$ & Pure $q$-anticoherent spin-$j$ state and its non-zero state multipoles (*) \tabularnewline
\hline 
\hline
$1$ & $1$ & 
$\begin{array}{l}
\rule[0pt]{0pt}{15pt}
|\psi\rangle=\frac{1}{\sqrt{2}}\left(\left|1,1\right\rangle +\left|1,-1\right\rangle \right)\\
\rho_{00} = \frac{1}{\sqrt{3}},\,
\rho_{20} =\frac{1}{\sqrt{6}},
\rho_{2-2} =\frac{1}{2}
\rule[-9pt]{0pt}{9pt}
\end{array}$
\tabularnewline
\hline 
$3/2$ & $1$ & 
$\begin{array}{l}
\rule[0pt]{0pt}{15pt}
|\psi\rangle=\frac{1}{\sqrt{2}}\left(\left|\tfrac{3}{2},\tfrac{3}{2}\right\rangle +\left|\tfrac{3}{2},-\tfrac{3}{2}\right\rangle \right)\\
\rho_{00} = \frac{1}{\sqrt{4}},\,
\rho_{20} =\frac{1}{2},
\rho_{3-3} =\frac{1}{2}
\rule[-9pt]{0pt}{9pt}
\end{array}$
\tabularnewline
\hline 
$2$ & $2$ & 
$\begin{array}{l}
\rule[0pt]{0pt}{15pt}
|\psi\rangle=\frac{1}{2}\left(\left|2,2\right\rangle +i\sqrt{2}\left|2,0\right\rangle +\left|2,-2\right\rangle \right)\\
\rho_{00} = \frac{1}{\sqrt{5}},\,
\rho_{3-2} =\frac{i}{2},
\rho_{4-4} =\frac{1}{4},
\rho_{40} =\frac{1}{2}\sqrt{\frac{7}{10}}
\rule[-9pt]{0pt}{9pt}
\end{array}$
\tabularnewline
\hline 
$5/2$ & $1$ & 
$\begin{array}{l}
\rule[0pt]{0pt}{15pt}
|\psi\rangle=\frac{1}{\sqrt{2}}\left(\left|\tfrac{5}{2},\tfrac{3}{2}\right\rangle +\left|\tfrac{5}{2},-\tfrac{3}{2}\right\rangle \right)\\
\rho_{00} = \frac{1}{\sqrt{6}},\,
\rho_{20} =-\frac{1}{2 \sqrt{21}},\,
\rho_{3-3} =\frac{1}{3}\,
\rho_{40} =-\frac{3}{2 \sqrt{7}},\,
\rho_{5-3} =-\frac{\sqrt{5}}{6}
\rule[-9pt]{0pt}{9pt}
\end{array}$
\tabularnewline
\hline 
$3$ & $3$ & 
$\begin{array}{l}
\rule[0pt]{0pt}{15pt}
|\psi\rangle=\frac{1}{\sqrt{2}}\left(\left|3,2\right\rangle +\left|3,-2\right\rangle \right)\\
\rho_{00} = \frac{1}{\sqrt{7}},\,
\rho_{4-4} =\frac{1}{2}\sqrt{\frac{5}{11}},\,
\rho_{40} =-\sqrt{\frac{7}{22}},\,
\rho_{6-4} =-\sqrt{\frac{3}{22}},\,
\rho_{60} =-\sqrt{\frac{3}{77}}
\rule[-9pt]{0pt}{9pt}
\end{array}$
\tabularnewline
\hline 
$7/2$ & $2$ & 
$\begin{array}{l}
\rule[0pt]{0pt}{15pt}
|\psi\rangle=\frac{\sqrt{2}}{3}\left|\tfrac{7}{2},\tfrac{7}{2}\right\rangle +\frac{\sqrt{\frac{7}{2}}}{3}\left(\left|\tfrac{7}{2},\tfrac{1}{2}\right\rangle +\left|\tfrac{7}{2},-\tfrac{5}{2}\right\rangle\right)
\\
\rho_{00} = \frac{1}{\sqrt{8}},\,
\rho_{3-3} =\frac{7}{3 \sqrt{66}},\,
\rho_{30} =\frac{7}{6 \sqrt{66}},\,
\rho_{5-3} =\frac{7}{6 \sqrt{78}},\,
\rho_{50} =\frac{7}{6} \sqrt{\frac{7}{78}},\,
\rho_{6-6} =\frac{1}{9}\sqrt{\frac{7}{2}},\,
\\
\rho_{6-3} =\frac{1}{18}\sqrt{\frac{77}{2}},\,
\rho_{60} =-\frac{1}{6}\sqrt{\frac{11}{6}},\,
\rho_{7-6} =\frac{1}{9}\sqrt{\frac{7}{2}},\,
\rho_{7-3} =-\frac{5}{18} \sqrt{\frac{35}{143}},\,
\rho_{70} =-\frac{8}{3} \sqrt{\frac{2}{429}}
\rule[-9pt]{0pt}{9pt}
\end{array}$
\tabularnewline
\hline 
$4$ & $3$ & 
$\begin{array}{l}
\rule[0pt]{0pt}{15pt}
|\psi\rangle=\frac{\sqrt{\frac{5}{6}}}{2}\left(\left|4,4\right\rangle +\left|4,-4\right\rangle \right)+\frac{\sqrt{\frac{7}{3}}}{2}\left|4,0\right\rangle \\
\rho_{00} = \frac{1}{3},\,
\rho_{4-4} = \frac{7}{6} \sqrt{\frac{5}{143}},\,
\rho_{40} =\frac{7}{3} \sqrt{\frac{7}{286}},\,
\rho_{6-4} =\frac{1}{3}\sqrt{\frac{35}{22}},\,
\rho_{60} =-\frac{1}{3}\sqrt{\frac{5}{11}},\,
\rho_{8-8} =\frac{5}{24},\,
\rho_{8-4} =\frac{1}{12}\sqrt{\frac{35}{13}},\,
\\
\rho_{80} =\frac{1}{4}\sqrt{\frac{55}{26}}
\rule[-9pt]{0pt}{9pt}
\end{array}$
\tabularnewline
\hline 
$9/2$ & $3$ & 
$\begin{array}{l}
\rule[0pt]{0pt}{15pt}
|\psi\rangle=\frac{1}{\sqrt{6}}\left(\left|\tfrac{9}{2},\tfrac{9}{2}\right\rangle +\left|\tfrac{9}{2},-\tfrac{9}{2}\right\rangle \right) -\frac{1}{\sqrt{3}} \left(\left|\tfrac{9}{2},\tfrac{3}{2}\right\rangle +\left|\tfrac{9}{2},-\tfrac{3}{2}\right\rangle \right) \\
\rho_{00} = \frac{1}{\sqrt{10}},\,
\rho_{3-3} = \frac{10 \sqrt{429}-3 \sqrt{2002}}{1287},\,
\rho_{40} =\frac{4}{\sqrt{715}},\,
\rho_{5-3} =-\frac{1}{117} \left(6 \sqrt{26}+\sqrt{273}\right),\,
\rho_{6-6} =-\frac{1}{\sqrt{30}},\,
\\
\rho_{60} =\frac{1}{2}\sqrt{\frac{5}{33}},\,
\rho_{7-3} =-\frac{5 \sqrt{3}-2 \sqrt{14}}{\sqrt{2431}},\,
\rho_{8-6} =-\frac{1}{3}\sqrt{\frac{7}{10}},\,
\rho_{80} =-\frac{1}{6}\sqrt{\frac{55}{13}},\,
\rho_{9-9} =\frac{1}{6},\,
\rho_{9-3} =-\frac{\sqrt{2}+2 \sqrt{21}}{3 \sqrt{221}}
\rule[-9pt]{0pt}{9pt}
\end{array}$
\tabularnewline
\hline 
$5$ & $3$ & 
$\begin{array}{l}
\rule[0pt]{0pt}{15pt}
|\psi\rangle=\frac{1}{\sqrt{11}}\left(\left|5,5\right\rangle +\left|5,-5\right\rangle \right)+\sqrt{\frac{3}{5}} \left|5,0\right\rangle\\
\rho_{00} = \frac{1}{\sqrt{11}},\,
\rho_{40} = 3 \sqrt{\frac{2}{143}},\,
\rho_{5-5} =\frac{3}{5 \sqrt{13}},\,
\rho_{60} =-3 \sqrt{\frac{3}{935}},\,
\rho_{7-5} =\frac{6}{\sqrt{221}},\,
\rho_{80} =2 \sqrt{\frac{22}{1235}},\,
\rho_{9-5} =\frac{1}{5}\sqrt{\frac{21}{17}},\,
\\
\rho_{10-10} =\frac{1}{5},\,
\rho_{100} =-\frac{29}{5} \sqrt{\frac{13}{3553}}
\rule[-9pt]{0pt}{9pt}
\end{array}$
\tabularnewline
\hline 
\end{tabular}
\caption{Examples of pure states with maximum order of anticoherence (\HOAP states) for the smallest values of the spin quantum number $j$, and their state multipoles. Here, $q$ is the order of anticoherence of the state. (*)~When a state multipole $\rho_{L -M}$ is non-zero, the same applies to $\rho_{LM}$ due to the relation $\rho_{LM}^*=(-1)^M\rho_{L-M}$. For the sake of simplicity, only the state multipoles $\rho_{L -M}$ with $M\geqslant 0$ are given in the table.
\label{tabACstates}}
\par\end{centering}
\end{table}

\clearpage

\end{widetext}

\appendix

\section{Master equation in the multipole operator basis}
In this appendix, we use the short-hand notation $\sum_{L,M}$ for $\sum_{L=0}^{N}\sum_{M=-L}^{L}$. Our aim is to write the master equation (\ref{MEq}) in the multipole operator basis (\basis) as defined below.

\subsection{Multipole operator basis (\basis)} \label{Appendix_MOB}
The multipole operator basis (\basis) is defined as the set $\{T_{LM}: L=0,\ldots, N; M=-L,\ldots,L \}$ of $(N+1)^2$ operators 
\begin{equation}
T_{LM}=\sqrt{\frac{2 L+1}{2j+1}} \sum_{m, m^{\prime}} \CG{j}{m}{L}{M}{j}{m'} \left|j, m\right\rangle\left\langle j, m^{\prime}\right|,
\end{equation}
where $N=2j$ and $\CG{j_1}{m_1}{j_2}{m_2}{j}{m}$ is the Clebsch-Gordan coefficient representing the probability amplitude that the coupling of the two angular momenta $j_{1}$ and $j_{2}$ with projections $m_{1}$ and $m_{2}$ results in the angular momentum $j$ with projection $m$. The operator set $\{T_{LM}\}$ forms an orthonormal basis of the space of linear operators acting on $\mathcal{H}\simeq\mathbb{C}^{N+1}$ verifying~\cite{Varshalovich}
\begin{equation}
\begin{aligned}
& T^{\dagger}_{LM}=\left(-1\right)^{M}T_{L-M},\\[3pt]
& \mathrm{Tr}\left[T_{LM}^{\dagger}T_{L'M'}\right]=\delta_{LL'}\delta_{MM'},\\[3pt]
& \mathrm{Tr}\left[T_{LM}\right]=\sqrt{N+1}\,\delta_{L0}\delta_{M0}.
\end{aligned}
\end{equation}
The multipole operators are irreducible tensor operators of rank $L$ and component $M$~\cite{Blum}.

We start with the expansion of the density operator in the \basis
\begin{equation}\label{rhoTLM}
\rho(t)=\sum_{L,M}\rho_{LM}(t)\,T_{LM}
\end{equation}
with the \emph{state multipoles}
\begin{equation}\label{rhoLMt}
\rho_{LM}(t)=\mathrm{Tr}\left[T_{LM}^{\dagger}\,\rho(t)\right]
\end{equation}
satisfying
\begin{equation}
\rho_{LM}^*=(-1)^M\rho_{L-M},\quad \rho_{00}=\frac{1}{\sqrt{N+1}}
\end{equation}
at all times because of Hermiticity and normalization of $\rho$. Additional nonlinear constraints on the components $\rho_{LM}$ follow from the positivity of $\rho$ (see, e.g., \cite{Byr03,Kry06}). In the \basis, the purity of a state $\rho$ reads as
\begin{equation}
\begin{aligned}
\mathrm{Tr}\left[\rho^{2}\right] & = \sum_{L,M}\sum_{L',M'}\rho_{LM}\,\rho_{L'M'}\mathrm{Tr}\left[T_{LM}T_{L'M'}\right]\\
& = \sum_{L,M}\left(-1\right)^{M}\,\rho_{LM}\,\rho_{L-M}\\
& = \sum_{L,M}\left|\rho_{LM}\right|^{2}.\label{purityrhoTLM}
\end{aligned}
\end{equation}
In particular, for pure states, $\sum_{L,M}\left|\rho_{LM}\right|^{2}=1$, and the $\left|\rho_{LM}\right|^{2}$ can be seen as forming a set of discrete probabilities.

The time derivative of $\rho_{LM}(t)$ defined in \eqref{rhoLMt} reads as
\begin{equation}
\begin{aligned}
\dot{\rho}_{LM}(t)=\frac{d}{dt}\left(\mathrm{Tr}\left[T_{LM}^{\dagger}\rho(t)\right]\right)=\mathrm{Tr}\left[T_{LM}^{\dagger}\dot{\rho}(t)\right]
\end{aligned}
\end{equation}
or, using the master equation $\dot{\rho}(t)=\mathcal{L}(\rho)$,
\begin{equation}\label{tdrhoLM}
\dot{\rho}_{LM}(t)=\mathrm{Tr}\left[T_{LM}^{\dagger}\mathcal{L}(\rho(t))\right]
\end{equation}

\subsection{Depolarization in terms of $\rho_{LM}$} \label{Appendix_MEMOB}

As the Lindblad operators $J_{\alpha}$ ($\alpha=x,y,z$) are Hermitian, the dissipator
\begin{equation}
\mathcal{D}_{\alpha}(\rho)=\gamma_{\alpha}\big(2J_{\alpha}\rho J_{\alpha}-J_{\alpha}J_{\alpha}\rho-\rho J_{\alpha}J_{\alpha}\big)
\end{equation}
can be rewritten as
\begin{equation}\label{Dalpha}
\mathcal{D}_{\alpha}(\rho)=-\gamma_{\alpha}\left[J_{\alpha},\left[J_{\alpha},\rho\right]\right].
\end{equation}
Let us first consider $\mathcal{D}_{z}(\rho)$. Using~\cite{Varshalovich}
\begin{equation}
\left[J_{z},T_{LM}\right]=\sqrt{L(L+1)}\,\CG{L}{M}{1}{0}{L}{M}T_{LM},
\end{equation}
we get
\begin{equation*}
\mathcal{D}_{z}(\rho)=-\gamma_{z}\sum_{L,M}\rho_{LM}\left[L\left(L+1\right)\right]
\left(\CG{L}{M}{1}{0}{L}{M}\right)^{2}
T_{LM}
\end{equation*}
From $\CG{L}{M}{1}{0}{L}{M}=M/\sqrt{L(L+1)}$, we conclude that the state multipoles $\rho_{LM}$ evolve under depolarization along $z$ according to
\begin{equation}\label{Dz}
\dot{\rho}_{LM}(t)|_{\mathcal{D}_{z}}\equiv 
\mathrm{Tr}\left[T_{LM}^{\dagger}\mathcal{D}_{z}(\rho)\right] =-\gamma_{z}\,M^{2}\rho_{LM}(t).
\end{equation}
Next, we consider $\mathcal{D}_{x}(\rho)$. By writing $J_{x}=(J_{+}+J_{-})/2$ and using the commutator~\cite{Varshalovich}
\begin{equation}\label{commJpmTLM}
\left[J_{\pm},T_{LM}\right]=\mp\sqrt{2L(L+1)}\,\CG{L}{M}{1}{\pm1}{L}{M\pm 1}T_{LM\pm 1}
\end{equation}
we get 
\begin{equation}
\begin{aligned}
\left[J_{x},\rho\right] ={}&	\frac{1}{2}\sum_{L,M}\rho_{LM}\left(\left[J_{-},T_{LM}\right]+\left[J_{+},T_{LM}\right]\right)\\
	={}&	\frac{1}{2}\sum_{L,M}\sqrt{2L(L+1)}\,\rho_{LM}\;\times\\
	&\quad \left(
	\CG{L}{M}{1}{-1}{L}{M-1} T_{LM-1} -\CG{L}{M}{1}{1}{L}{M+1} T_{LM+1}\right)
\end{aligned}
\end{equation}
and therefore
\begin{widetext}
\begin{equation}
\begin{aligned}
\left[J_{x},\left[J_{x},\rho\right]\right]	=	\frac{1}{2}\sum_{L,M}L(L+1)\,\rho_{LM}\Big[ & \CG{L}{M}{1}{-1}{L}{M-1}\left(\CG{L}{M-1}{1}{-1}{L}{M-2}\,T_{LM-2}-\CG{L}{M-1}{1}{1}{L}{M}\,T_{LM}\right)\\
& -\CG{L}{M}{1}{1}{L}{M+1}\left(\CG{L}{M+1}{1}{-1}{L}{M}\,T_{LM}-\CG{L}{M+1}{1}{1}{L}{M+2}\,T_{LM+2}\right)\Big].
\end{aligned}
\end{equation}
Using the explicit formula for Clebsch-Gordan coefficients~\cite{Varshalovich}
\begin{equation}
\CG{a}{\alpha}{b}{\beta}{a+b-1}{\alpha+\beta}=2\,(b\alpha-a\beta)\left(\frac{(2a+2b-1)\,(2a-1)!\,(2b-1)!\,(a+b+\alpha+\beta-1)!\,(a+b-\alpha-\beta-1)!}{(a+\alpha)!\,(a-\alpha)!\,(b+\beta)!\,(b-\beta)!\,(2a+2b)!}\right)^{1/2}
\end{equation}
we eventually get
\begin{align}
& \dot{\rho}_{LM}(t)|_{\mathcal{D}_{x}}= -\frac{\gamma_{x}}{4}\Big[2\left(L^{2}+L-M^{2}\right)\rho_{LM} +d_{LM}^{+}\, \rho_{LM+2}+d_{LM}^{-}\, \rho_{LM-2}\Big]\label{Dx}
\end{align}
where
\begin{equation}
d_{LM}^{\pm} = \sqrt{(L\mp M)(L\pm M+1)(L\mp M-1)(L\pm M+2)}.
\end{equation}
A similar calculation for $J_y=(J_{+}-J_{-})/(2i)$ yields
\begin{align}
 & \dot{\rho}_{LM}(t)|_{\mathcal{D}_{y}} = -\frac{\gamma_{y}}{4}\Big[2\left(L^{2}+L-M^{2}\right)\rho_{LM} -d_{LM}^{+}\,\rho_{LM+2}-d_{LM}^{-}\,\rho_{LM-2}\Big].\label{Dy}
\end{align}
\end{widetext}
Eventually, by adding all three contributions \eqref{Dz}, \eqref{Dx}, and \eqref{Dy}, we get Eq.~\eqref{eq:ME_TLM}.

\section{Models of depolarization} \label{Appendix_models}

In this appendix, we present models that provide a physical basis for the master equation (\ref{MEq}). A first model, inspired from~\cite{2009Muller}, is that of a spin system interacting with a fluctuating magnetic field $\mathbf{B}(t)$. The interaction Hamiltonian is
\begin{equation}
	H_{\mathrm{int}} = -\mu\, \mathbf{J}\cdot\mathbf{B},
\end{equation}
where $\mathbf{J}=(J_x,J_y,J_z)$ and $\boldsymbol{\mu}=\mu \mathbf{J}$ is the spin magnetic moment. We choose a coarse-graining time $\Delta t$ for the evolution and make the assumption that this time is much larger than the correlation time $\tau_c$ of the fluctuations of the magnetic field. Then, assuming white noise and isotropic magnetic field fluctuations, the two-time correlation functions of the components of $\mathbf{B}(t)$ are given by
\begin{equation}
	\overline{B_\alpha(t_1)B_\beta(t_2)}=\frac{B^2}{3}\tau_c\,\delta(t_1-t_2)\,\delta_{\alpha\beta},
\end{equation}
where $\alpha,\beta=(x,y,z)$ and an overbar denotes the ensemble average over realizations of the stochastic process. Finally, if we make the Born approximation that no correlations appear between the system and the magnetic field, by taking the limit $\Delta t \to dt$, it can be shown that the evolution reduces to
\begin{equation}
	\frac{d\rho}{dt}=\sum_{\alpha=x,y,z}\gamma\big[J_\alpha,[J_\alpha,\rho]\big],
\end{equation}
where $\gamma=\tau_c\omega_0^2/2$ is the effective Larmor frequency $\omega_0=\mu B/(\sqrt{3}\hbar)$. The limit $\Delta t \to dt$ is valid as long as $\Delta t\ll \tau_s$ where $\tau_s=1/\gamma$ is the typical evolution time of the system, hence, $\tau_c \ll \Delta t \ll \tau_s$.

As a second model, we consider a collection of two-level atoms of electric dipole moment $\mathbf{d}$ interacting collectively with the electromagnetic field at thermal equilibrium. This system is governed by the usual superradiant master equation~\cite{Aga74}
\begin{equation}\label{optical_master_equation}
	\begin{aligned}
		\frac{d\rho}{dt} ={}& \gamma_0(\bar{n}+1)\Big[2J_-\rho J_+ - \big(J_+J_-\rho + \rho J_+J_-\big)\Big] \\
											& +\gamma_0\bar{n}\Big[2J_+\rho J_- - \big(J_-J_+\rho + \rho J_-J_+\big)\Big],
	\end{aligned}
\end{equation}
where $\gamma_0 = 4\omega^3|\mathbf{d}|^2/3\hbar c^3$ is the spontaneous emission rate and $\bar{n}=(e^{\hbar \omega / k_B T}-1)^{-1}$ is the average number of thermal photons at the atomic transition frequency $\omega$. The first line of Eq.~\eqref{optical_master_equation} describes spontaneous and thermally induced emission of photons while the second line describes the absorption of thermal photons. When $k_BT \gg \hbar\omega$, the average number of photons $\overline{n}\gg 1$ and $\overline{n}+1\approx \overline{n}$, so that the rates for emission and absorption are roughly the same. In this case, the master equation (\ref{optical_master_equation}) reduces to
\begin{equation}
	\frac{d\rho}{dt} = -\gamma_{\perp z}\Big(\big[J_x,[J_x,\rho]\big]+\big[J_y,[J_y,\rho]\big]\Big)
\end{equation}
with $\gamma_{\perp z} = 2\gamma_0\overline{n}$ and thus provides a model for anisotropic depolarization.

Finally, a third model is based on a weak continuous measurement of the spin components. When an observable $A$ of a quantum system is continuously measured with an apparatus and that the measurement is not read out, the system evolves according to~\cite{2006Jacobs}
\begin{equation}\label{cont_meas}
	\frac{d\rho}{dt}=-\gamma\big[A,[A,\rho]\big]
\end{equation}
with $\gamma>0$ a constant proportional to the strength of the measurement. We immediately recognize the form of our dissipator (\ref{Dissalpha}). Thus, in order to obtain the full master equation (\ref{MEq}), the three collective spin observables $J_x$, $J_y$, and $J_z$ must be simultaneously subjected to a continuous measurement. Because these observables are not compatible, cross terms of the form $\big[J_\alpha,[J_\beta,\rho]\big]$ with $\alpha\ne\beta$ appear in addition to the simple sum of terms of the form (\ref{cont_meas}). However, these extra terms can be neglected within the Born-Markov approximation~\cite{2020Jiang}.

\section{Partial trace in the \basis} \label{Appendix_MOBptrace}

In this Appendix, we show that if $\rho$ is an $N$-qubit symmetric state, with expansion in the \basis~given by
\begin{equation}\label{rhoTLMappendix}
\rho=\sum_{L=0}^{N}\sum_{M=-L}^{L}\rho_{LM}T_{LM}^{(N)},
\end{equation}
then its $q$-qubit reduction $\rho_q\equiv\mathrm{Tr}_{\neg q}[\rho]$ is given by
\begin{equation}\label{rhotTLM}
\rho_q=\sum_{L=0}^{q}\sum_{M=-L}^{L} \tfrac{q!}{N!}\sqrt{\tfrac{(N-L)!(N+L+1)!}{(q-L)!(q+L+1)!} }\rho_{LM}T_{LM}^{(q)},
\end{equation}
where, for clarity, we have added a superscript to the multipolar operators  indicating the number of qubits. To prove the above result, we work in the spin-coherent state representation (or symmetric separable state representation). Denoting by $|\Omega\rangle$ the state of a qubit with Bloch vector pointing in the direction of the unit vector $\mathbf{n}=(\sin\theta\cos\varphi,\sin\theta\sin\varphi,\cos\theta)^T$ and $d\Omega=\sin\theta d\theta d\varphi$, the multipole operators read as \cite{Gil76}
\begin{equation}\label{TLMPrep}
T^{(N)}_{LM}=\tfrac{N+1}{4 \pi}\alpha_{L}^{(N)} \int Y_{LM}(\Omega) \left(|\Omega\rangle\langle \Omega|\right)^{\otimes N} d \Omega,
\end{equation} 
with $Y_{LM}(\Omega)$ the spherical harmonics and $\alpha_{L}^{(N)}$ the constant
\begin{equation}
\alpha_{L}^{(N)}=2\sqrt{\pi}\,\tfrac{\sqrt{(N-L)!(N+L+1)!}}{(N+1)!}.
\end{equation}
In the coherent state representation, the partial trace is readily performed because the partial trace of $\left(|\Omega\rangle\langle\Omega|\right)^{\otimes N}$ over any set of $N-q$ qubits is trivially $\left(|\Omega\rangle\langle \Omega|\right)^{\otimes q}$. Thus, the partial trace of \eqref{TLMPrep} yields
\begin{equation}
\begin{aligned}
\mathrm{Tr}_{\neg q}\left[T_{LM}^{(N)}\right]={}& \tfrac{N+1}{4 \pi} \alpha_{L}^{(N)}\int Y_{LM}(\Omega) \left(|\Omega\rangle\langle \Omega|\right)^{\otimes q} d \Omega
\end{aligned}
\end{equation}
or, upon comparison with \eqref{TLMPrep} where $N$ has been replaced by $q$,
\begin{equation}\label{ptraceTLM}
\mathrm{Tr}_{\neg q}\left[T_{LM}^{(N)}\right]=
\left\{\begin{array}{ll}
\frac{N+1}{q+1}\frac{\alpha_{L}^{(q)}}{\alpha_{L}^{(N)}}\,T_{LM}^{(q)} & L\leqslant q, \\[10pt]
0 &L>q.
\end{array}\right.
\end{equation}
Equation \eqref{ptraceTLM} for $L>q$ holds true because the integral
\begin{equation}\label{TLMPrep"}
\int Y_{LM}(\Omega)\left(|\Omega\rangle\langle \Omega|\right)^{\otimes q} d \Omega
\end{equation}
vanishes for $L>q$, as can be seen by substituting Eq.~\eqref{cohTLM} for $\left(|\Omega\rangle\langle \Omega|\right)^{\otimes q}$ and using the orthogonality of  spherical harmonics. In \eqref{ptraceTLM}, the prefactor of $T_{LM}^{(q)} $ can also be written as
\begin{equation}\label{alpharel}
\tfrac{N+1}{q+1}\tfrac{\alpha_{L}^{(q)}}{\alpha_{L}^{(N)}}=\tfrac{q!}{N!}\sqrt{\tfrac{(N-L)!(N+L+1)!}{(q-L)!(q+L+1)!}}.
\end{equation}
In the end, by linearity of the partial trace, using Eqs.~\eqref{rhoTLMappendix}, \eqref{ptraceTLM}, and \eqref{alpharel}, we get \eqref{rhotTLM}. A similar result to the one presented here was obtained in~\cite{Dev07} by a different method.

\subsection*{Purity of reduced density matrices} \label{Appendix_reducedpurity}

The purity of the $q$-qubit reduced density operator \eqref{rhotTLM} of a state $\rho$ with state multipoles $\rho_{LM}$ is given by
\begin{equation}\label{purityrhotTLM}
\tr\left[\rho_q^{2}\right]=\sum_{L=0}^{q}\sum_{M=-L}^{L} \tfrac{(q!)^2}{(N!)^2}\tfrac{(N-L)!(N+L+1)!}{(q-L)!(q+L+1)!}\, |\rho_{LM}|^2.
\end{equation}
In particular, for $q=N-1$, we have
\begin{equation}\label{purityrhoNm1TLM0}
\mathrm{Tr}\left[\rho_{N-1}^{2}\right]=\sum_{L=0}^{N-1}\sum_{M=-L}^{L} \tfrac{N(N+1)-L (L+1)}{N^2}\, |\rho_{LM}|^2.
\end{equation}
Then, by combining Eq.~\eqref{purityrhoTLM} for $\mathrm{Tr}\left[\rho^{2}\right]$ with Eq.~\eqref{purityrhoNm1TLM0}, and after some algebra, we arrive at the identity
\begin{equation}\label{purityrhoNm1TLM}
\begin{aligned}
& \sum_{L=0}^{N}\sum_{M=-L}^{L} L (L+1)\, |\rho_{LM}|^2\\
& \quad = N(N+1)\,\mathrm{Tr}\left[\rho^{2}\right] - N^2\,\mathrm{Tr}\left[\rho_{N-1}^{2}\right]
\end{aligned}
\end{equation}
which is valid for any state $\rho$ (pure or mixed). As the measure of anticoherence $\mathcal{A}_{q}$ is a function of $\tr\left[\rho_q^{2}\right]$, it can be expressed in terms of state multipoles using~\eqref{purityrhotTLM}.

\section{Evolution of purities} \label{Appendix_purities}

In this appendix, we give analytical solutions to the system of equations \eqref{setequR} for the purities, which we reproduce below:
\begin{equation}\label{syseqRAppendix}
\begin{aligned}
& \dpuritydt(\rho_{1}) = -2\gamma\left[2\purity(\rho_{1}) - 1\right], \\
& \dpuritydt(\rho_{2}) = -2\gamma\left[6\purity(\rho_{2}) - 4\purity(\rho_{1})\right],\\
& \qquad\vdots \\
& \dpuritydt(\rho_{N-1}) = -2\gamma[(N-1)N\purity(\rho_{N-1}) \\
& \phantom{\dpuritydt(\rho_{N-1}) = -2\gamma [(N-1)N}- (N-1)^2\purity(\rho_{N-2})],\\
& \dpuritydt(\rho) = -2\gamma\left[N(N+1)\purity(\rho) - N^2\purity(\rho_{N-1})\right].
\end{aligned}
\end{equation}
We start by noting that the first $q<N$ equations of \eqref{syseqRAppendix} also form a closed set of equations. Therefore, the time evolution of $\purity(\rho_q)$ depends only on the initial purities $R_q(0),R_{q-1}(0),\ldots,R_1(0)$, where we have used the shorthand notation $R_q(0)$ for $\purity(\rho_{q}(0))$. The general solutions are then given, for $q=1$, by
\begin{equation*}\label{R1t}
R_{1}(t) = \frac{e^{-4 \gamma t}}{2} \big[ 2\, R_{1}(0)-1\big]+\frac{1}{2},
\end{equation*}
for $q=2$, by
\begin{equation*}\label{R2t}
\begin{aligned}
R_{2}(t) ={}& \frac{e^{-12 \gamma  t}}{6}  \Big[ 2\, e^{12 \gamma  t} +3\,(2\, R_{1}(0) - 1)\, e^{8 \gamma  t} \\
& \phantom{\frac{e^{-12 \gamma  t}}{6}  \Big[}+6\, \big(R_{2}(0) - R_{1}(0)\big) +1\Big],
\end{aligned}
\end{equation*}
and, for $q=3$, by
\begin{equation*}\label{R3t}
\begin{aligned}
R_3(t) ={}& \frac{e^{-24 \gamma  t}}{20}  \Big[ e^{12 \gamma  t} \Big(9\, (2 R_1(0)-1) \, e^{8 \gamma  t}\\ 
& -30 R_1(0) +30 R_2(0)+5 \, e^{12 \gamma  t}+5\Big)\\
& +12 R_1(0) -30 R_2(0) +20 R_3(0)-1\Big].
\end{aligned}
\end{equation*}
The system of equations \eqref{syseqRAppendix}, supplemented with the dummy variable $R_0=1$ and the corresponding equation $\dot{R}_0=0$, can be written in matrix form as $\dot{\mathbf{R}}^T=M\mathbf{R}^T$ with $\mathbf{R}=(R(0),R_{N-1}(0),\ldots,R_1(0),R_0(0))$ and $M$ a matrix with non zero entries only on the main diagonal and the subdiagonal. The eigenvalues of $M$ are therefore given by its diagonal elements $-2\gamma q (q+1)$ with $q=1,\ldots,N$. This explains the decay rates appearing in the three expressions above. 

Furthermore, for pure initial states, it holds that $R_q(0)=R_{N-q}(0)$ and the solution for $R(\rho)$ can be expressed solely in terms of the initial purities $R_{\lfloor N/2\rfloor}(0),\ldots,R_{1}(0)$ or, equivalently, in terms of anticoherence measures \eqref{ACR}. 

\section{Preservation of separability under isotropic depolarization} \label{Appendix_coherencepreservation}

In this Appendix, we show that the dynamics generated by the master equation~\eqref{MEq} for $\gamma_{x}=\gamma_{y}=\gamma_z$ preserves the separability of states, i.e., that a separable state can only evolve into a separable state under isotropic depolarization. As the Hamiltonian part only induces a rotation in the state space, it clearly preserves the separability and coherence of a state. For notational convenience, we set $\hbar=1$ in the following. We start by rewriting the dissipative part of the master equation as
\begin{equation}\label{MEA}
	\sum_{\alpha=x,y,z}\mathcal{D}_{\alpha}(\rho) = \mathcal{D}_{+z}(\rho) + \mathcal{D}_{-z}(\rho),
\end{equation}
where
\begin{equation*}
		\mathcal{D}_{\pm z}(\rho) = \gamma \Big( J_\pm\rho J_\mp - \frac{1}{2}\{ J_\mp J_\pm , \rho \} + J_z\rho J_z - \frac{1}{2}\{ J_z^2 , \rho \}\Big).
\end{equation*}
We then follow the method of Gisin~\cite{1992Gisin} to show that certain quantum diffusion equations for spin relaxation preserve spin-coherent states.
This method is based on the unravelling of the master equation into stochastic quantum trajectories and will allow us to prove that both dissipators $\mathcal{D}_{\pm z}$ are separability preserving.

First, consider as in~\cite{1992Gisin} the deterministic equation for the relaxation of a spin
\begin{equation}
\label{eq:deterministic_relaxation_z}
	\frac{d|\psi\rangle}{dt}=\left(\langle J_z \rangle - J_z \right)|\psi\rangle
\end{equation}
If the initial state $|\psi\rangle$ is a coherent state (or separable symmetric state in the language of multiqubit systems), then at any later time the state will also be coherent (and therefore separable). This is because a state is coherent iff $|\langle \mathbf{J}\rangle|^2  \equiv \sum_{i=x,y,z} \langle J_i \rangle ^2  = j^2$, and because $|\langle \mathbf{J}\rangle|^2 $ is a quantity conserved under spin relaxation for a coherent initial state. Indeed, it holds that
\begin{equation}
	\frac{d}{dt} |\langle \mathbf{J}\rangle|^2(t)  = 2\sum_{i=x,y,z} \langle J_i\rangle \left(2\langle J_i\rangle\langle J_z\rangle - \langle \{J_i,J_z\} \rangle\right)
\end{equation}
For a coherent state initially pointing in the direction of the unit vector $\mathbf{n}$, we have $\langle J_k\rangle = j n_k$ and $\left\langle\{J_k,J_\ell\}\right\rangle=  j(2j+1)n_k n_\ell + j\delta_{k\ell}$, from which follows that 
\begin{equation}
	\frac{d}{dt} |\langle \mathbf{J}\rangle|^2(0) = 2j^2\sum_{i=x,y,z} n_k(n_k n_z - \delta_{kz}) = 0
\end{equation}
As the direction $z$ in Eq.~\eqref{eq:deterministic_relaxation_z} is arbitrary, the result is equally valid for $J_x$ and $J_y$. Moreover, a similar calculation shows that it is also valid for jump operators $J_\pm$. The next step is to write a stochastic differential equation (SDE) for the master equation \eqref{MEA}. The Stratonovitch SDE associated with $\mathcal{D}_{\pm z}$ reads as~\cite{1998Diosi}
\begin{equation}\label{SDEStrato}
\begin{aligned}
	 {}& d|\psi\rangle= \sqrt{\gamma}\big[(J_\pm - \langle J_\pm\rangle) + (J_z - \langle J_z\rangle) \big] |\psi\rangle d\xi \\ 
	& + \gamma \Big[\langle J_\mp \rangle (J_\pm -\langle J_\pm \rangle) - \frac{1}{2} J_\mp  J_\pm + \frac{1}{2} \langle J_\mp  J_\pm \rangle\Big]|\psi\rangle dt \\
	& + \gamma \Big[ \langle J_z\rangle (J_z - \langle J_z \rangle) - \frac{1}{2} J_z^2 + \frac{1}{2} \langle J_z^2 \rangle\Big]|\psi\rangle dt,
\end{aligned}
\end{equation}
where the first line is the diffusion term and the second and third lines are the drift terms.
Then, due to the relation $J_\mp J_\pm - J_z^2 = j(j+1) \mp J_z$, we can write
\begin{equation*}
\begin{aligned}
& \langle J_- \rangle J_+ - \langle J_- \rangle \langle J_+ \rangle = \langle J_- \rangle (J_+ - \langle J_+ \rangle),\\
& \langle J_-J_+\rangle + \langle J_z^2 \rangle - J_-J_+ - J_z^2 = J_z - \langle J_z \rangle
\end{aligned}
\end{equation*}
so that Eq.~\eqref{SDEStrato} can now be written as a sum of terms of the form (\ref{eq:deterministic_relaxation_z}) for the jump operators $J_z$ and $J_+$. Both preserve spin-coherent states, which eventually shows that a coherent state remains coherent when subjected to isotropic depolarization.

\section{Positivity of the $P$ function}
\label{Appendix_Pfunc}

The set of coherent states forms an overcomplete basis in which any state $\rho$ can be expanded. This leads to the $P$-representation of $\rho$, which for multiqubit symmetric density matrices reads as
\begin{equation}\label{rhoPfunc}
\rho=\frac{N+1}{4 \pi} \int P_{\rho}(\Omega)\,\left(|\Omega\rangle\langle \Omega|\right)^{\otimes N} d \Omega.
\end{equation}
Here, the $P$-function $P_{\rho}(\Omega)$ is a non-unique real quasiprobability distribution over $S^2$. When $P_{\rho}(\Omega)$ is non-negative for all $\Omega\equiv(\theta,\varphi)$, the state \eqref{rhoPfunc} is separable because it is a convex combination of separable symmetric states. The $P$ function being an angular function, it can always be expanded over the spherical harmonics. By truncating the expansion to keep only spherical harmonics with orbital quantum numbers $L\leqslant N$, we end up with an equally valid $P$ function for $\rho$~\cite{2008Giraud}:
\begin{equation}\label{PYlm}
	\tilde{P}_{\rho}(\Omega)  =\sum_{L,M} P_{LM}\, Y_{LM}(\Omega),
\end{equation}
where~\cite{1981Agarwal}
\begin{equation}
	P_{LM} = \tfrac{(-1)^{L-M}}{N!\sqrt{4\pi}}\sqrt{(N-L)!(N+L+1)!}\,\rho_{LM}
\end{equation}
with $\rho_{LM}$ the state multipoles of $\rho$. Using $Y_{00} = 1/\sqrt{4\pi}$, $\rho_{00}=1/\sqrt{N+1}$, $\rho_{L-M} = (-1)^M\, \rho^*_{LM}$, $Y_{L-M} = (-1)^M \, Y^*_{LM}$ and the solution \eqref{gensoldiag} for the matrix elements in the case of isotropic depolarization, Eq.~\eqref{PYlm} yields
\begin{align}
		\tilde{P}_{\rho(t)}(\Omega) & = \frac{1}{4\pi} + \sum_{L=1}^N \frac{\beta(N,L)}{N!} \, e^{-L(L+1)t} \times\label{Pfunctime}\\
& \sum_{M=0}^L (-1)^{M} (2-\delta_{M0})\, \mathrm{Re}\big[Y_{LM}(\Omega) e^{-i\omega Mt} \rho_{LM}(0)\big]\nonumber
\end{align}
with $\beta(N,L)= (-1)^{L}\sqrt{(N-L)!(N+L+1)!/(4\pi)}$. For an initial entangled state $\rho$, the corresponding $P$ function, given by Eq.~\eqref{Pfunctime} with $t=0$, is not everywhere positive. However, after some time, the exponentially decaying term in the right-hand side of Eq.~\eqref{Pfunctime} becomes small enough to make the $P$ function positive for all $\Omega$. We define $t_P$ as the minimum time after which this occurs, ensuring that the state \eqref{rhoPfunc} is a convex combination of separable states, hence is separable at $t=t_P$. This also implies that the state is separable for all $t\geqslant t_P$ because the dynamics is separability preserving. 

To estimate $t_P$ numerically, we computed a discretized version of the $P$ function \eqref{Pfunctime} for $\Omega=(\theta,\varphi)\in [0,\pi]\times [0,2\pi]$ sampled on a grid with resolution $\delta\theta=\delta\varphi=1/N^{3/4}$. Then we checked at what time, with precision $\delta t=1/(4096 \sqrt{N})$, this discretized $P$ function becomes everywhere positive.

\section{Decoherence rates for pure states} \label{Appendix_PureDecoRate}

In this appendix, we use the short-hand notation $\sum_{L,M}$ for $\sum_{L=0}^{N}\sum_{M=-L}^{L}$. Our aim is to evaluate, for \emph{pure states}, the decoherence rates of isotropic and anisotropic depolarization appearing in Eqs.~\eqref{Rdotisotropic} and \eqref{Rdotanisotropic}. Up to a constant factor, they are given by
\begin{equation*}
\begin{aligned}
& \sum_{L,M} L (L+1)\, |\rho_{LM}|^2,\\
& \sum_{L,M} M^2\, |\rho_{LM}|^2,\\
& \sum_{L,M} \left[L (L+1)-M^2\right]\, |\rho_{LM}|^2.
\end{aligned}
\end{equation*}
We make use of the fact that
\begin{equation}\label{purestate}
\rho=|\psi\rangle\langle\psi|\;\: \Rightarrow \;\: \left\{
\begin{array}{l}
\mathrm{Tr}\left[\rho^{2}\right]=1,\\[4pt]
\mathrm{Tr}\left[\rho_{N-1}^{2}\right]=\mathrm{Tr}\left[\rho_{1}^{2}\right].
\end{array}\right.
\end{equation}
For $\rho=|\psi\rangle\langle\psi|$, Eq.~\eqref{purityrhoNm1TLM} can then be simplified with the help of Eq.~\eqref{purestate} into 
\begin{equation}\label{purityrhoNm1TLMpure}
\sum_{L,M} L (L+1)\, |\rho_{LM}|^2 = N(N+1)- N^2\mathrm{Tr}\left[\rho_{1}^{2}\right].
\end{equation}
Now, it holds that for any state $\rho$ (pure or mixed), $\mathrm{Tr}\left[\rho_{1}^{2}\right]$ can be expressed as a function of the sum of the variances $\Delta J_\alpha^2= \langle J_\alpha^2 \rangle-\langle J_\alpha \rangle^2$ of the spin components $J_\alpha$ ($\alpha=x,y,z$) in state $\rho$, a quantity also known as the total variance~\cite{Kly03}
\begin{equation}\label{eq:TVspin}
\begin{aligned}
\mathbb{V} &\equiv \sum_{\alpha=x,y,z} \Delta J_\alpha^2 = j(j+1)-|\langle \mathbf{J}\rangle|^2,
\end{aligned}
\end{equation}
where $j=N/2$.
The total variance quantifies the overall level of quantum fluctuations of the spin. It is minimal for spin-coherent states and maximal when $\langle \mathbf{J}\rangle=0$ (in particular, this holds for pure anticoherent states and for the maximally mixed state). In terms of total variance, we have~\cite{Bag17}
\begin{equation}\label{purityrho1TV}
\mathrm{Tr}\left[\rho_{1}^{2}\right]=1-\frac{\mathbb{V}-j}{2j^2}=\frac{N(N+1)-2\,\mathbb{V}}{N^2}.
\end{equation}
This relationship allows us to rewrite the first decoherence rate \eqref{purityrhoNm1TLMpure} as
\begin{equation}\label{purityrhoNm1TLMpurefinal}
\begin{aligned}
\sum_{L,M} L (L+1)\, |\rho_{LM}|^2 ={}& 2\left(\Delta J_x^2+\Delta J_y^2+\Delta J_z^2\right) \\
 ={}& N\left(\frac{N}{2}+1\right)- 2\,|\langle \mathbf{J}\rangle|^2.
\end{aligned}
\end{equation}

We then show that the second decoherence rate is given, for pure states, by
\begin{equation}\label{dr2pure}
\sum_{L,M} M^2\, |\rho_{LM}|^2 = 2\, \Delta J_z^2.
\end{equation}
The left-hand side of Eq.~\eqref{dr2pure} can be obtained by combining the commutator $[J_z,T_{LM}]=M\,T_{LM}$ with the expansion \eqref{rhoTLMappendix} of the density operator and the orthonormality relation $\mathrm{Tr}\left[T_{LM}^{\dagger}T_{L'M'}\right]=\delta_{LL'}\delta_{MM'}$ of the \basis as
\begin{equation}\label{dr2purea}
	\mathrm{Tr}\left([J_z,\rho][\rho,J_z]\right)=\sum_{L,M} M^2|\rho_{LM}|^2.
\end{equation}
For a pure state $\rho=|\psi\rangle\langle\psi|$, the left-hand side of \eqref{dr2purea} is equal to $2\Delta J_z^2$. Indeed, by expanding the commutators, we get
\begin{equation}
	\mathrm{Tr}\left([J_z,\rho][\rho,J_z]\right)=2\left[\mathrm{Tr}(J_z^2\rho^2)-\mathrm{Tr}(J_z\rho J_z\rho)\right].
\end{equation}
Using \eqref{purestate}, the first term is given by
\begin{equation}
\begin{aligned}
\mathrm{Tr}(J_z^2\rho^2) & = \mathrm{Tr}(J_z^2\rho) = \langle J_z^2\rangle
\end{aligned}
\end{equation}
while the second term is given by
\begin{equation}
\begin{aligned}
 \mathrm{Tr}(J_z\rho J_z\rho) & = \mathrm{Tr}\left(J_z|\psi\rangle\langle\psi|J_z|\psi\rangle\langle\psi|\right) \\
 & = \langle\psi|J_z|\psi\rangle\langle\psi|J_z|\psi\rangle = \langle J_z\rangle^2.
\end{aligned}
\end{equation}
Eventually, for pure states, it holds that $\mathrm{Tr}\left([J_z,\rho][\rho,J_z]\right)=2(\langle J_z^2\rangle-\langle J_z\rangle^2)=2\Delta J_z^2$.

Last, by substracting Eq.~\eqref{dr2pure} from Eq.~\eqref{purityrhoNm1TLMpurefinal}, we readily get the third decoherence rate 
\begin{equation*}
\sum_{L,M} \left[L(L+1)-M^2\right]\, |\rho_{LM}|^2 = 2\, (\Delta J_x^2+\Delta J_y^2).
\end{equation*}


\begin{thebibliography}{99}
\bibitem{2018Zeilinger} M.\ Erhard, R.\ Fickler, M.\ Krenn and A. Zeilinger, Twisted photons: new quantum perspectives in high dimensions, \href{https://doi.org/10.1038/lsa.2017.146}{Light: Science \& Applications \textbf{7}, 17146 (2018).}

\bibitem{2021Satoor} T.\ Satoor, A.\ Fabre, J.-B.\ Bouhiron, A.\ Evrard, R.\ Lopes, and S.\ Nascimbene, Partitioning dysprosium's electronic spin to reveal entanglement in non-classical states, \href{https://arxiv.org/abs/2104.14389} {arXiv:2104.14389.}

\bibitem{Ionqudits} P.~J.~Low, B.~M.~White, A.~A.~Cox, M.~L.~Day, and C.~Senko, Practical trapped-ion protocols for universal qudit-based quantum computing, \href{https://doi.org/10.1103/PhysRevResearch.2.033128}{Phys.\ Rev.\ Research \textbf{2}, 033128 (2020).}

\bibitem{SCqudits} M.\ Neeley, M.\ Ansmann, R.\ C.\ Bialczak, M.\ Hofheinz, E.\ Lucero, A.\ D.\ O'Connell, D.\ Sank, H.\ Wang, J.\ Wenner, A.\ N.\ Cleland, M.\ R.\ Geller, J.\ M.\ Martinis, Emulation of a Quantum Spin with a Superconducting Phase Qudit, \href{https://doi.org/10.1126/science.1173440}{Science \textbf{325}, 722 (2009).}

\bibitem{Saw20} R.\ Sawant, J.\ A.\ Blackmore, P.\ D.\ Gregory, J.\ Mur-Petit, D.\ Jaksch, J.\ Aldegunde, J.\ M.\ Hutson, M.\ R.\ Tarbutt and S.\ L.\ Cornish, Ultracold polar molecules as qudits, \href{https://doi.org/10.1088/1367-2630/ab60f4}{New J.~Phys.~\textbf{22}, 013027 (2020).}

\bibitem{2002Kempe} C.\ Simon and J.\ Kempe, Robustness of multiparty entanglement, \href{https://doi.org/10.1103/PhysRevA.65.052327}{Phys.\ Rev.\ A \textbf{65}, 052327 (2002).}

\bibitem{2004Dur} W.\ Dür and H.-J.\ Briegel, Stability of Macroscopic Entanglement under Decoherence, \href{https://doi.org/10.1103/PhysRevLett.92.180403}{Phys.\ Rev.\ Lett.\ \textbf{92}, 180403 (2004).}

\bibitem{2005Dur} M.\ Hein, W.\ Dür and H.-J.\ Briegel, Entanglement properties of multipartite entangled states under the influence of decoherence, \href{https://doi.org/10.1103/PhysRevA.71.032350}{Phys.\ Rev.\ A \textbf{71}, 032350 (2005).}

\bibitem{2005Lidar} S.\ Bandyopadhyay and D.\ Lidar, Robustness of multiqubit entanglement in the independent decoherence model, \href{https://doi.org/10.1103/PhysRevA.72.042339}{Phys.\ Rev.\ A \textbf{72}, 042339 (2005).}

\bibitem{Aol08} L.\ Aolita, R.\ Chaves, D.\ Cavalcanti, A.\ Acín and L.\ Davidovich, Scaling Laws for the Decay of Multiqubit Entanglement, \href{https://doi.org/10.1103/PhysRevLett.100.080501}{Phys.\ Rev.\ Lett.\ \textbf{100}, 080501 (2008).}

\bibitem{2010Nori} X.\ Wang, A.\ Miranowicz, Y.\ Liu, C.\ P.\ Sun and F.\ Nori, Sudden vanishing of spin squeezing under decoherence, \href{https://doi.org/10.1103/PhysRevA.81.022106}{Phys.\ Rev.\ A \textbf{81}, 022106 (2010).}

\bibitem{2013Rivas} Á.\ Rivas and A.\ Luis, SU($2$)-invariant depolarization of quantum states of light, \href{https://doi.org/10.1103/PhysRevA.88.052120}{Phys.\ Rev.\ A \textbf{88}, 052120 (2013).}

\bibitem{Benedict99} M.\ G.\ Benedict and A.\ Czirjáks, Wigner functions,
squeezing properties, and slow decoherence of a meso-
scopic superposition of two-level atoms, \href{https://doi.org/10.1103/PhysRevA.60.4034}{Phys.\ Rev.\ A \textbf{60}, 4034 (1999).}

\bibitem{Gol18} A.\ Z.\ Goldberg and D.\ F.\ V.\ James, Quantum-limited Euler angle measurements using anticoherent states, \href{https://doi.org/10.1103/PhysRevA.98.032113}{Phys.\ Rev.\ A \textbf{98}, 032113 (2018).}

\bibitem{Mar20} J.\ Martin, S.\ Weigert and O.\ Giraud, Optimal detection of rotations about unknown axes by coherent and anticoherent states, \href{https://doi.org/10.22331/q-2020-06-22-285}{Quantum \textbf{4} 285 (2020).}

\bibitem{Gol21} A.\ Z.\ Goldberg, A.\ B.\ Klimov, G.\ Leuchs and L.\ L.\ Sánchez-Soto, Rotation sensing at the ultimate limit, \href{https://doi.org/10.1088/2515-7647/abeb54}{J.\ Phys.\ Photonics \textbf{3}, 022008 (2021).}

\bibitem{2013Kosloff} R.\ Kosloff, Quantum Thermodynamics: A Dynamical Viewpoint, \href{https://doi.org/10.3390/e15062100}{Entropy \textbf{15}, 2100 (2013).}

\bibitem{2020Kosloff} R.\ Dann and R.\ Kosloff, Open system dynamics from thermodynamic compatibility, \href{https://doi.org/10.1103/PhysRevResearch.3.023006}{Phys.\ Rev.\ Research \textbf{3}, 023006 (2021).}

\bibitem{Arecchi72} F.\ T.\ Arecchi, E.\ Courtens, R.\ Gilmore, ad H.\ Thomas, Atomic Coherent States in Quantum Optics, \href{https://doi.org/10.1103/PhysRevA.6.2211}{Phys.\ Rev.\ A \textbf{6}, 2211 (1972).}

\bibitem{2017Giraud} F.\ Bohnet-Waldraff, O.\ Giraud, and D.\ Braun, Absolutely classical spin states, \href{https://doi.org/10.1103/PhysRevA.95.012318}{Phys.\ Rev.\ A \textbf{95}, 012318 (2017).}

\bibitem{2008Giraud} O.\ Giraud, P.\ Braun, and D.\ Braun, Classicality of spin states, \href{https://dx.doi.org/10.1103/PhysRevA.78.042112}{Phys. Rev. A. \textbf{78}, 042112 (2008).}

\bibitem{2006Zimba} J.\ Zimba, ``Anticoherent” Spin States via the Majorana Representation, \href{http://www.ejtp.com/articles/ejtpv3i10p143.pdf}{Electr.\ J.\ Theor.\ Phys.\ \textbf{3}, 143 (2006).}

\bibitem{Gir15}  O.\ Giraud, D.\ Braun, D.\ Baguette, T.\ Bastin, and J.\ Martin, Tensor Representation of Spin States, \href{https://doi.org/10.1103/PhysRevLett.114.080401}{Phys.~Rev.~Lett.~\textbf{114}, 080401 (2015).}

\bibitem{Bag15} D.\ Baguette, F.\ Damanet, O.\ Giraud, and J.\ Martin, Anticoherence of spin states with point-group symmetries, \href{https://doi.org/10.1103/PhysRevA.92.052333}{Phys.~Rev.~A \textbf{92}, 052333 (2015).}

\bibitem{ACstates} \href{http://www.oq.ulg.ac.be/libspinACstates.html}{http://www.oq.ulg.ac.be/libspinACstates.html}, \href{http://polarization.markus-grassl.de/}{http://polarization.markus-grassl.de/}.

\bibitem{Bag17} D.\ Baguette and J.\ Martin, Anticoherence measures for pure spin states, \href{https://doi.org/10.1103/PhysRevA.96.032304}{Phys.~Rev.~A \textbf{96}, 032304 (2017).}

\bibitem{2005Korbicz} J.\ K.\ Korbicz, J.\ I.\ Cirac and M.\ Lewenstein, Spin Squeezing Inequalities and Entanglement of $N$ Qubits States, \href{https://doi.org/10.1103/PhysRevLett.95.120502}{Phys.\ Rev.\ Lett.\ \textbf{95}, 120502 (2005).}

\bibitem{Zyc01} M.\ Kuś and K.\ Życzkowski, Geometry of entangled states, \href{https://doi.org/10.1103/PhysRevA.63.032307}{Phys.\ Rev.\ A \textbf{63}, 032307 (2001).}

\bibitem{Bjo15} G.\ Björk, A.\ B.\ Klimov, P.\ de la Hoz, M.\ Grassl, G.\ Leuchs, and L.\ L.\ Sánchez-Soto, Extremal quantum states and their Majorana constellations, \href{https://doi.org/10.1103/PhysRevA.92.052333}{Phys.~Rev.~A \textbf{92}, 031801(R) (2015).}

\bibitem{2014Rozema} L.\ A.\ Rozema, D.\ H.\ Mahler, R.\ Blume-Kohout and A.\ Steinberg, Optimizing the Choice of Spin-Squeezed States for Detecting and Characterizing Quantum Processes, \href{https://doi.org/10.1103/PhysRevX.4.041025}{Phys.\ Rev.\ X \textbf{4}, 041025 (2014).}

\bibitem{Tidstr2003} J.\ Tidström and E.\ Sjöqvist, Uhlmann's geometric phase in presence of isotropic decoherence \href{https://doi.org/10.1103/PhysRevA.67.032110}{Phys.\ Rev.\ A \textbf{67}, 032110 (2003).}

\bibitem{2006Klimov} A.\ B.\ Klimov, J.\ L.\ Romero and L.\ L.\ Sánchez Soto, Single quantum model for light depolarization, \href{https://doi.org/10.1364/JOSAB.23.000126}{J.\ Opt.\ Soc.\ Am.\ B \textbf{23}, 126 (2006).}

\bibitem{2017Arsenijevic} M.\ Arsenijević, J.\ Jeknić-Dugić and M.\ Dugić, Generalized Kraus Operators for the One-Qubit Depolarizing Quantum Channel, \href{https://doi.org/10.1007/s13538-017-0502-3}{Braz.\ J.\ Phys.\ \textbf{47}, 339 (2017).}

\bibitem{2021Chen} H-B.\ Chen, Effects of symmetry breaking of the structurally-disordered Hamiltonian ensembles on the anisotropic decoherence, \href{https://arxiv.org/abs/2104.13237}{arXiv:2104.13237.}

\bibitem{2017Bhattacharya} S.\ Bhattacharya, A.\ Misra, C.\ Mukhopadhyay and A.\ K.\ Pati, Exact master equation for a spin interacting with a spin bath: Non-Markovianity and negative entropy production rate, \href{https://doi.org/10.1103/PhysRevA.95.012122}{Phys.\ Rev.\ A \textbf{95}, 012122 (2017).}

\bibitem{2009Muller} C.\ A.\ Müller, \href{https://doi.org/10.1007/978-3-540-88169-8_6}{\emph{Diffusive Spin Transport}} in A.\ Buchleitner, C.\ Viviescas and M.\ Tiersch, \emph{Entanglement and Decoherence}, \href{https://doi.org/10.1007/978-3-540-88169-8}{Lecture Notes in Physics 768, Springer-Verlag Berlin Heidelberg 2009.}

\bibitem{BreuerPetruccione} H.\ P.\ Breuer and F.\ Petruccione, \emph{The theory of open quantum systems}, \href{https://doi.org/10.1093/acprof:oso/9780199213900.001.0001}{Oxford University Press, 2002.}

\bibitem{Lid05} D.\ A.\ Lidar, A.\ Shabani, R.\ Alicki, Conditions for strictly purity-decreasing quantum Markovian dynamics, \href{https://doi.org/10.1016/j.chemphys.2005.06.038}{Chemical Physics \textbf{322}, 82 (2006).}

\bibitem{Louck} L.\ C.\ Biedenharn and J.\ D.\ Louck, \emph{Angular Momentum in Quantum Physics Theory and Application}, \href{https://doi.org/10.1017/CBO9780511759888}{Cambridge University Press, 1984.}

\bibitem{Varshalovich} D.\ A.\ Varshalovich, A.\ N.\ Moskalev and V.\ K.\ Khersonskii, \emph{Quantum Theory of Angular Momentum}, \href{https://doi.org/10.1142/0270}{(World Scientific Publishing Company, Singapore, 1988).}

\bibitem{Hor09} K.\ Hornberger, \href{https://doi.org/10.1007/978-3-540-88169-8_5}{\emph{Introduction to Decoherence Theory}} in A.\ Buchleitner, C.\ Viviescas and M.\ Tiersch, \href{https://doi.org/10.1007/978-3-540-88169-8}{\emph{Entanglement and Decoherence}, Lecture Notes in Physics 768, Springer-Verlag Berlin Heidelberg 2009.}

\bibitem{Nielsen_Chuang} M.\ A.\ Nielsen, I.\ L.\ Chuang, \emph{Quantum Computation and Quantum Information} \href{https://doi.org/10.1017/CBO9780511976667}{Cambridge University Press, Cambridge, 2000.}

\bibitem{2019Kattem} J.\ Kattemölle and J.\ van Wezel, Conditions for superdecoherence, \href{https://doi.org/10.22331/q-2020-05-14-265}{Quantum \textbf{4}, 265 (2020).}

\bibitem{2005Berman} G.\ P.\ Berman, D.\ I.\ Kamenev, and V.\ I.\ Tsifrinovich, Collective decoherence of the superpositional entangled states in the quantum Shor algorithm, \href{https://doi.org/10.1103/PhysRevA.71.032346}{Phys.\ Rev.\ A \textbf{71}, 032346 (2005).}

\bibitem{2011Monz} T.\ Monz, P.\ Schindler, J.\ T.\ Barreiro, M.\ Chwalla, D.\ Nigg, W.\ A.\ Coish, M.\ Harlander, W.\ Hänsel, M.\ Hennrich, and R.\ Blatt, 14-Qubit Entanglement : Creation and Coherence, \href{https://doi.org/10.1103/PhysRevLett.106.130506}{Phys.\ Rev.\ Lett.\ \textbf{106}, 130506 (2011).}

\bibitem{footnoteRenyi} The Rényi entropies are defined as $S_{q}(\rho) = \frac{1}{1-q} \ln \left(\tr[\rho^{q}]\right)$ for $q>0$. For $q=2$, $S_{2}(\rho)=-\ln \left(\tr[\rho^{2}]\right)=-\ln \left(R(\rho)\right)$ and we have $S_{2}(\rho) \geq S_{2}\left(\rho_{t}\right)\;\Leftrightarrow\; R(\rho)\leqslant R(\rho_t)$.

\bibitem{Zyczkowski_book} I.\ Bengtsson and K.\ {\.Z}yczkowski, \emph{Geometry of Quantum States : An Introduction to Quantum Entanglement}, \href{https://doi.org/10.1017/9781139207010}{2nd ed.\ Cambridge University Press 2017.}

\bibitem{Wol14} E.\ Wolfe and S.\ F.\ Yelin, Certifying Separability in Symmetric Mixed States of $N$ Qubits, and Superradiance, \href{https://doi.org/10.1103/PhysRevLett.112.140402}{Phys.~Rev.~Lett.~\textbf{112}, 140402 (2014).}

\bibitem{2019Campaioli} F.\ Campaioli, F.\ A.\ Pollock and K.\ Modi, Tight, robust, and feasible quantum speed limits for open dynamics, \href{https://doi.org/10.22331/q-2019-08-05-168}{Quantum \textbf{3}, 168 (2019).}

\bibitem{2016Uzdin} R.\ Uzdin and R.\ Kosloff, Speed limits in Liouville space for open quantum systems, \href{https://doi.org/10.1209/0295-5075/115/40003}{EPL \textbf{115}, 40003 (2016).}

\bibitem{spectralnorm} G.\ Strang, \emph{Linear Algebra and Its Applications}, 4th ed.\ New York: Academic Press, 1980. 

\bibitem{Tot09} G.\ Toth and O.\ Gühne, Entanglement and Permutational Symmetry, \href{https://doi.org/10.1103/PhysRevLett.102.170503}{Phys.\ Rev.\ Lett.\ \textbf{102}, 170503 (2009).}

\bibitem{Tot10} G.\ Toth and O.\ Gühne, Separability criteria and entanglement witnesses for symmetric quantum states, \href{https://doi.org/10.1103/PhysRevLett.102.170503}{Appl.\ Phys.\ B \textbf{98}, 617 (2010).}

\bibitem{Eck02}  K.\ Eckert, J.\ Schliemann, D.\ Bru{\ss}, and M.\ Lewenstein, Quantum Correlations in Systems of Indistinguishable Particles, \href{https://doi.org/10.1006/aphy.2002.6268}{Ann.~Phys.~\textbf{299}, 88 (2002).}

\bibitem{1998Horodecki} M.\ Horodecki, P.\ Horodecki and R.\ Horodecki, Mixed-State Entanglement and Distillation:  Is there a “Bound” Entanglement in Nature?, \href{https://doi.org/10.1103/PhysRevLett.80.5239}{Phys.\ Rev.\ Lett.\ \textbf{80}, 5239 (1998).}

\bibitem{Nat18} N.\ Johnston and E.\ Patterson, The inverse eigenvalue problem for entanglement witnesses, \href{https://doi.org/10.1016/j.laa.2018.03.043}{Linear Algebra and its Applications \textbf{550}, 1–27 (2018).}

\bibitem{Vid02} G.~Vidal and R.~F.~Werner, Computable measure of entanglement, \href{https://doi.org/10.1103/PhysRevA.65.032314}{Phys.\ Rev.\ A \textbf{65}, 032314 (2002).}

\bibitem{2003Stockton} J.\ K.\ Stockton, J.\ M.\ Geremia, A.\ C.\ Doherty, and H.\ Mabuchi, Characterizing the entanglement of symmetric many-particle spin-$\tfrac{1}{2}$ systems, \href{https://doi.org/10.1103/PhysRevA.67.022112}{Phys.\ Rev.\ A \textbf{67}, 022112 (2003).}

\bibitem{footnotenegcomp} To compute the negativity for large numbers of qubits, we have adapted the function \textsf{sym2bipartite} from QUBIT4MATLAB V5.5~\cite{QUBIT4MATLAB} to the Julia programming language.

\bibitem{QUBIT4MATLAB} G.\ Toth, QUBIT4MATLAB V3.0: A program package for quantum information science and quantum optics for MATLAB, \href{https://doi.org/10.1016/j.cpc.2008.03.007}{Comput.\ Phys.\ Comm.\ \textbf{179}, 430 (2008). }

\bibitem{footnotetNPT} To accurately estimate $t_\mathrm{NPT}$, the moment when the negativity becomes zero, we had to work with octuple precision floats. We worked with the BigFloat type available in \href{https://julialang.org}{Julia} for floating point numbers of arbitrary precision, which uses the \href{https://www.mpfr.org}{GNU MPFR} library.

\bibitem{2015Aolita} L.\ Aolita, F.\ de Melo and L.\ Davidovich, Open-system dynamics of entanglement:a key issues review, \href{https://doi.org/10.1088/0034-4885/78/4/042001}{Rep.\ Prog.\ Phys.\ \textbf{78}, 042001 (2015).}

\bibitem{BE5o2} More precisely, the time-evolved state is NPT with respect to the bipartitions $(2,3)$ and $(1,4)$ for $\gamma t$ up to $0.127961$ and $0.129957$ respectively.  

\bibitem{BE3} More precisely, the time-evolved state is NPT with respect to the bipartitions $(3,3)$, $(2,4)$ and $(1,5)$ for $\gamma t$ up to $0.111822$, $0.107694$, $0.1149$ respectively.  

\bibitem{1996Palma} G.\ M.\ Palma, K.-A.\ Suominen and A.\ K.\ Ekert, Quantum computers and dissipation, \href{https://doi.org/10.1098/rspa.1996.0029}{Proc.\ R.\ Soc.\ Lond.\ A \textbf{452}, 567 (1996).}

\bibitem{2002Reina} J.\ H.\ Reina, L.\ Quiroga and N.\ F.\ Johnson, Decoherence of quantum registers, \href{https://doi.org/10.1103/PhysRevA.65.032326}{Phys.\ Rev.\ A \textbf{65}, 032326 (2002).}

\bibitem{Li07} S.-B.\ Li, J.-B.\ Xu, Robust and fragile Werner states in the collective dephasing, \href{https://doi.org/10.1140/epjd/e2006-00216-x}{Eur.\ Phys.\ J.\ D \textbf{41}, 377 (2007).}

\bibitem{2008Guhne} O.\ Gühne, F.\ Bodoky, and M.\ Blaauboer, Multiparticle entanglement under the influence of decoherence, \href{https://doi.org/10.1103/PhysRevA.78.060301}{Phys.\ Rev.\ A \textbf{78}, 060301(R) (2008).}

\bibitem{2019Korbicz} J.\ Tuziemski, A.\ Lampo, M.\ Lewenstein and J.\ K.\ Korbicz, Reexamination of the decoherence of spin registers, \href{https://doi.org/10.1103/PhysRevA.99.022122}{Phys.\ Rev.\ A \textbf{99}, 022122 (2019).}

\bibitem{2003Lidar} D.\ A.\ Lidar and K.\ B.\ Whaley, Decoherence-Free Subspaces and Subsystems . In: F.\ Benatti, R.\ Floreanini (eds) \emph{Irreversible Quantum Dynamics}. \href{https://doi.org/10.1007/3-540-44874-8_5}{Lecture Notes in Physics, \textbf{622}. Springer, Berlin, Heidelberg (2003).}

\bibitem{Ser04} A.\ Serafini, S.\ De Siena, F.\ Illuminati, and M.\ G.\ A.\ Paris, Minimum decoherence cat-like states in Gaussian noisy channels, \href{https://doi.org/10.1088/1464-4266/6/6/019}{J.\ Opt.\ B: Quantum Semiclass.\ Opt.\ \textbf{6}, S591 (2004).}

\bibitem{2018Jeannic} H.\ Le Jeannic, A.\ Cavaillès, K.\ Huang, R.\ Filip, and J.\ Laurat, Slowing Quantum Decoherence by Squeezing in Phase Space, \href{https://doi.org/10.1103/PhysRevLett.120.073603}{Phys.\ Rev.\ Lett.\ \textbf{120}, 073603 (2018).}

\bibitem{2018Brewster} R.\ A.\ Brewster, T.\ B.\ Pittman, and J.\ D.\ Franson, Reduced decoherence using squeezing, amplification, and antisqueezing, \href{https://doi.org/10.1103/PhysRevA.98.033818}{Phys.\ Rev.\ A \textbf{98}, 033818 (2018).}

\bibitem{2020Gebia}  F.\ Gebbia, C.\ Benedetti, F.\ Benatti, R.\ Floreanini, M.\ Bina, and M.\ G.\ A.\ Paris, Two-qubit quantum probes for the temperature of an Ohmic environment, \href{https://doi.org/10.1103/PhysRevA.101.032112}{Phys.\ Rev.\ A \textbf{101}, 032112 (2020).}

\bibitem{Kuk21}  S.\ Kukita, Y.\ Matsuzaki, Y.\ Kondo, Heisenberg-limited quantum metrology using collective dephasing, \href{https://arxiv.org/abs/2103.11612}{arXiv:2103.11612.}

\bibitem{Rac17} C.\ Rackauckas and Q.\ Nie, DifferentialEquations.jl – A Performant and Feature-Rich Ecosystem for Solving Differential Equations in Julia, \href{https://doi.org/10.5334/jors.151}{Journal of Open Research Software 5(1), 15 (2017).}

\bibitem{Blum} K.\ Blum, \emph{Density Matrix Theory and Applications}, \href{https://doi.org/10.1007/978-3-642-20561-3}{Springer (2011).}

\bibitem{Byr03} M.\ S.\ Byrd and N.\ Khaneja, Characterization of the positivity of the density matrix in terms of the coherence vector representation, \href{https://doi.org/10.1103/PhysRevA.68.062322}{Phys.\ Rev.\ A \textbf{68}, 062322 (2003).}

\bibitem{Kry06} S.\ Kryszewski and M.\ Zachciał, Positivity of the $N \times N$ density matrix expressed in terms of polarization operators, \href{https://doi.org/10.1088/0305-4470/39/20/019}{J.\ Phys.\ A: Math.\ Gen.\ \textbf{39}, 5921 (2006).}

\bibitem{Aga74} G.\ S.\ Agarwal, \emph{Quantum statistical theories of spontaneous emission and their relation to other approaches}, in Höhler G. (eds) \href{https://doi.org/10.1007/BFb0042382}{Quantum Optics. Springer Tracts In Modern Physics \textbf{70}, 1 (1974).}

\bibitem{2006Jacobs} K.\ Jacobs and D.\ A.\ Steck, A Straightforward Introduction to Continuous Quantum Measurement, \href{https://doi.org/10.1080/00107510601101934}{Contemporary Physics \textbf{47}, 279 (2006).}

\bibitem{2020Jiang} C.\ Jiang and G.\ Watanabe, Quantum dynamics under simultaneous and continuous measurement of noncommutative observables, \href{https://doi.org/10.1103/PhysRevA.102.062216}{Phys.\ Rev.\ A \textbf{102}, 062216 (2020).}

\bibitem{Gil76}  R.\ Gilmore, Q and P representatives for spherical tensors, \href{https://doi.org/10.1088/0305-4470/9/7/001 }{J.\ Phys.\ A \textbf{9}, L65 (1976).}

\bibitem{Dev07} A.\ R.\ Usha Devi, R.\ Prabhu, and A.\ K.\ Rajagopal, Collective multipolelike signatures of entanglement in symmetric $N$-qubit systems, \href{https://doi.org/10.1103/PhysRevA.76.012322}{Phys.\ Rev.\ A \textbf{76}, 012322 (2007).}

\bibitem{1992Gisin} N.\ Gisin and M.\ B.\ Cibils, Quantum diffusions, quantum dissipation and spin relaxation, \href{https://doi.org/10.1088/0305-4470/25/19/024}{J.\ Phys.\ A: Math.\ Gen.\ \textbf{25}, 5165 (1992).}

\bibitem{1998Diosi}  L.\ Diosi, N.\ Gisin and W.\ T.\ Strunz, Non-Markovian quantum state diffusion, \href{https://doi.org/10.1103/PhysRevA.58.1699}{Phys.\ Rev.\ A \textbf{58}, 1699 (1998).}

\bibitem{1981Agarwal} G.\ S.\ Agarwal, Relation between atomic coherent-state representation, state multipoles, and generalized phase-space distributions, \href{https://doi.org/10.1103/PhysRevA.24.2889}{Phys.\ Rev.\ A \textbf{24}, 2889 (1981).}

\bibitem{Kly03}  A.\ A.\ Klyachko and A.\ S.\ Shumovsky, Entanglement, local measurements and symmetry, \href{https://doi.org/10.1088/1464-4266/5/3/364}{J.\ Opt.\ B: Quantum Semiclass.\ Opt.\ \textbf{5}, S322 (2003).}

\end{thebibliography}
\end{document}